\documentclass[fleqn,usenatbib]{mnras}
\usepackage{Preamble}

\usepackage{newtxtext,newtxmath}

\usepackage[T1]{fontenc}

\DeclareRobustCommand{\VAN}[3]{#2}
\let\VANthebibliography\thebibliography
\def\thebibliography{\DeclareRobustCommand{\VAN}[3]{##3}\VANthebibliography}

\usepackage{graphicx}	
\usepackage{amsmath}	

\title[Extreme broad He II emission at high and low z and VMS]{Extreme broad He\2 emission at high and low redshifts: the dominant role of VMS in NGC 3125-A1 and CDFS131717.}

\author[A. Wofford et al.]{\parbox{\textwidth}{\thanks{E-mail: awofford @ astro.unam.mx}
Aida Wofford$^{1}$\orcidlink{0000-0001-8289-3428},
Andrés Sixtos$^{1}$\orcidlink{0000-0001-6452-2797},
Stephane Charlot$^{2}$\orcidlink{0000-0003-3458-2275},
Gustavo Bruzual$^{3}$\orcidlink{0000-0002-6971-5755},
Fergus Cullen$^{4}$\orcidlink{0000-0002-3736-476X},
Thomas M. Stanton$^{4}$\orcidlink{0000-0002-0827-9769},
Svea Hernández$^{5}$\orcidlink{0000-0003-4857-8699},
Linda J. Smith$^{6}$\orcidlink{0000-0002-0806-168X},
Matthew Hayes$^{7}$\orcidlink{0000-0001-8587-218X}}
\\\\
$^{1}$Instituto de Astronom\'ia, Universidad Nacional Aut\'onoma de M\'exico, Unidad Acad\'emica en Ensenada, Km 103 Carr. Tijuana$-$Ensenada, Ensenada, B.C.,\\C.P. 22860, M\'exico
\\
\parbox{\textwidth}{$^{2}$Sorbonne Universit\'e, CNRS, UMR 7095, Institut d'Astrophysique de Paris, 98 bis bd Arago, 75014 Paris, France}\\
\parbox{\textwidth}{$^{3}$Instituto de Radioastronomía y Astrofísica, UNAM Campus Morelia, Apartado postal 3-72, 58090 Morelia, Michoacán, México}\\
\parbox{\textwidth}{$^{4}$Institute for Astronomy, University of Edinburgh, Royal Observatory, Edinburgh, EH9 3HJ, UK}\\
\parbox{\textwidth}{$^{5}$AURA for ESA, Space Telescope Science Institute, 3700 San Martin Drive, Baltimore, MD 21218, USA}\\
\parbox{\textwidth}{$^6$Space Telescope Science Institute, 3700 San Martin Drive, Baltimore, MD 21218, USA}\\
$^7$Stockholm University, Department of Astronomy and Oskar Klein Centre for Cosmoparticle Physics, AlbaNova University Centre, SE-10691, Stockholm, Sweden
}
\date{Accepted XXX. Received YYY; in original form ZZZ}

\pubyear{2015}

\newcommand{\onlyinsubfile}[1]{#1}

\begin{document}

\renewcommand{\onlyinsubfile}[1]{}

\label{firstpage}
\pagerange{\pageref{firstpage}--\pageref{lastpage}}
\maketitle


\begin{abstract}

Super star cluster (SSC) A1 ($3.1\times10^5\,M_\odot$) in NGC~3125 has one of the strongest ($EW=4.6\pm0.5\,$\AA) broad ($FWHM=1131\pm40\,$km\,s$^{-1}$) He\2~$\lambda$1640 emission lines in the nearby Universe and constitutes an important template for interpreting observations of extreme He\2 emitters out to redshifts of $z\sim2-3$. We use Cosmic Origins Spectrograph (COS) observations of A1 to show that there is no significant contamination of the He\2 line with nebular emission and that the line is redshifted by $121\pm17$\,km\,s$^{-1}$ relative to ISM lines. We compare  the COS G130M + G160M observations of A1 to recent binary BPASS and single-star Charlot \& Bruzual (C\&B) simple stellar population (SSP) models with Very Massive Stars (VMS) of up to $300\,M_\odot$. We suggest why BPASS models fail to reproduce A1's He\2 emission. On the other hand, a  C\&B model with $Z=0.008$, age = 2.2 Myr, and VMS approaching the Eddington limit provides an excellent fit to the He\2~emission and fits reasonably well C\3~$\lambda$1175, N\5~$\lambda\lambda1238,1241$, and C\4~$\lambda\lambda$1548, 1551. We present O\5~$\lambda$1371 line-profile predictions showing that this line constitutes an important tracer of youth and VMS in galaxies. Finally, we discuss the presence of VMS in CDFS131717, a highly star-forming low-metallicity galaxy located at $z=3.071$, which has a tentative detection of O\5 absorption and strong broad He\2 emission. These features are rare and hint to the presence of short-lived VMS in the galaxy. Our results show the effect of the latest developments of stellar wind theory and the importance of accounting for VMS in models.

\end{abstract} 



\section{Introduction}\label{sec:1}

\subsection{Starburst galaxies, massive stars, and cosmic noon}

Starburst galaxies are characterised by massive violent bursts of star formation. They are powered by populations of hot stars with progenitor masses of 5 M$_\odot$ and above, which have main diagnostic lines in the satellite ($<3200$\,\AA) ultraviolet \citep[UV][]{2001ApJ...550..724L}. Locally, starburst galaxies are responsible for 20\% of  star formation in galaxies \citep{2004MNRAS.351.1151B}. Nearby starburst galaxies are the most obvious local counterparts of the normal star-forming galaxies discovered at high redshift, in particular galaxies located at redshifts of $z\sim2-3$, i.e., at the epoch of `cosmic noon'. The epoch of cosmic noon is ideal to examine the mechanisms of star formation because it is when galaxies formed about half of their current stellar mass \citep{2020ARA&A..58..661F}. Stacked spectra of star-forming cosmic-noon galaxies \citep[e.g.,][]{2003ApJ...588...65S, 2019MNRAS.487.2038C, 2020A&A...636A..47S}, as well as single-object observations \citep{2000ApJ...528...96P,2010ApJ...719.1168E,2019Sci...366..738R,2020MNRAS.499L..67V}  provide a glimpse of the dominant stellar populations present at the peak of star formation in the Universe, and in particular, the massive-star ($\ge10\,M_\odot$) content. 

\subsection{The broad He\2 problem at high and low redshift}\label{sec:he2pb}

 Broad He\2~$\lambda$1640 emission (hereafter, He\2~1640) is the strongest stellar line in the stacked UV spectrum of 811 Lyman Break Galaxies (LBGs) located at $z\sim3$ produced by \cite{2003ApJ...588...65S}.  Lyman Break Galaxies are star-forming galaxies that are selected using their differing appearance in several imaging filters due to the position of the Lyman limit. When \cite{2003ApJ...588...65S} published their composite spectrum, the He\2 line was the only strongly discrepant stellar line when comparing the rest-frame UV spectra of LBGs to the UV templates of local starbursts, in particular those available at the time \citep{1999ApJS..123....3L} from the widely used stellar population synthesis package Starburst99 \citep{2010ApJS..189..309L}. 

In stacked spectra, the uncertainties in the redshifts of the galaxies that make up the spectrum contribute to some extent to the width of the composite He\2~line profile. However, a recent deep (20 hr) {\it Very Large Telescope} (\vlt) observation of a UV bright (M$_{\rm UV}$ = -21.7) non-lensed star-forming galaxy (ID = CDFS131717) that is located at $z_{\rm spec}=3.071$ and is part of the  deep public ESO spectroscopic survey with VIMOS on the \vlt, VANDELS \citep{2018MNRAS.479...25M,2021A&A...647A.150G}, clearly shows the existence of broad  He\2~1640 emission in single objects (Stanton et al., in prep.). 

On the other hand, some nearby starburst galaxies show extreme broad He\2 emission lines that until very recently were also challenging to model with population synthesis models. This is the case of SSC A1 in galaxy NGC~3125. The extreme He\2 $\lambda4686$ emission of this SSC was first noted by \cite{1981A&A...101L...5K}. In their  study of nearby starburst galaxies with WR-star features observed with \hst's Space Telescope Imaging Spectrograph (STIS), \cite{2004ApJ...604..153C} confirmed the extreme nature of A1 using the He\2 1640 line, which is the UV equivalent of the optical line at 4686 \AA. The latter authors also studied NGC~3125 in the context of LBGs located at $z\sim3$.  Finally, \cite{2021MNRAS.503.6112S} studied a different sample of nearby broad He\2 1640 emitters observed with \hst's Cosmic Origins Spectrograph (COS, an instrument with higher sensitivity and spectral resolution with respect to the STIS / G140L setup used by \citealt{2004ApJ...604..153C}). The later authors only found one star-forming galaxy, SB 179, with a strength of the He\2 1640 emission line  similar to that of \nai.

In integrated spectra of massive-star populations, broad ($FWHM\approx1000$\,km\,s$^{-1}$) He\2~ emission suggests stellar winds as the formation mechanism. Strong, broad He\2~1640 emission requires a high mass-loss rate (to produce a dense wind) and high temperature (to doubly ionise He), with He not necessarily overabundant (though He\2 emission is boosted if the star is He rich).
 
Since classical Wolf-Rayet (cWR) stars (He-burning, H-deficient) meet the first two conditions and are He rich, the presence of broad He\2~is generally thought to indicate the presence of WR stars. On the other hand, Very Massive Stars (VMS), which have initial masses of $M\geq100\,$M$_\odot$ \citep{2015ASSL..412.....V}, i.e., above the commonly-used upper mass limit of the stellar IMF, can produce WR features even during core H-burning, independent of He enrichment. These are WNh stars, where the "W" stands for "Wolf-Rayet" and the "N" and "h" indicate the dominance of nitrogen emission lines and presence of hydrogen lines in the star's spectrum, respectively {\bf\citep{2008ApJ...679.1467S}}. In VMS, the reason for the increased $EW$ of the He\2~$\lambda4686$ stellar-wind line is almost certainly the proximity to the Eddington limit. This is shown in figure 10 of \citet{2011A&A...531A.132V}, where one sees that increasing the electron-scattering Eddington factor theoretically increases the mass loss rate, and also empirically, the He\2~ $EW$. A similar increase of the $EW$ is expected for the UV line as well. 

VMS have been found in star cluster R136, which is at the center of LMC's large H\2 region NGC 2070 \citep{2010MNRAS.408..731C, 2016MNRAS.458..624C}. R136 is $1-2.5\,$Myr old \citep{2022A&A...663A..36B} and has a binary-corrected virial mass of $M=4.6-14.2\times10^4\,M_\odot$ \citep{2012A&A...546A..73H} that is consistent with the photometric mass of $\sim5\times10^4\,M_\odot$ that is reported in \cite{2009ApJ...707.1347A}. There are several  determinations of the initial mass of the most massive star in R136, R136a1: $320^{+100}_{-40}\,M_\odot$ \citep[based on \hst~UV spectroscopy from the Goddard High Resolution Spectrograph and the Faint Object Spectrograph $+$ optical \vlt~/ SINFONI spectroscopy $+$ \vlt~/ MAD IR, according to their table 1]{2010MNRAS.408..731C},
 $251^{+48}_{-31}$ \citep[based on STIS UV spectroscopy]{2020MNRAS.499.1918B}, $273^{+25}_{-36}$ $M_\odot$ \citep[based on STIS optical and UV spectroscopy]{2022A&A...663A..36B} and $196^{+34}_{-27}\,M_\odot$ \cite[based on BVRI-like photometry from the Gemini speckle imager Zorro, which separates sources a1 and \#9]{2022ApJ...935..162K}. On the other hand, in W14, we ruled out that the extraordinary He\2~emission and O\5~$\lambda$1371 (O\5~1371) absorption of \nai\,are due to an extremely flat exponent of the upper IMF, and suggested that they originate in the winds of VMS. Furthermore, \cite{2016ApJ...823...38S} found evidence of the presence of VMS in cluster \#5 of blue compact dwarf (BCD) galaxy NGC 5253 based on \hst~/ UV and \vlt~/ optical spectra, and \cite{2018ApJ...865...55L} suggested that VMS could explain the COS far-UV observations of SSC-N in BCD galaxy II~Zw~40.

Several research groups have developed population synthesis models, i.e., models that predict the photometric and spectroscopic properties of stellar systems, which account for the contribution of VMS  \citep[e.g.][]{2016MNRAS.462.1757G, 2022MNRAS.512.5329B, 2022A&A...659A.163M}. In particular, \cite{2022A&A...659A.163M} discuss the expected effects of VMS in integrated spectra of massive-star populations and galaxies that host them. \cite{2022A&A...659A.163M} were the first to successfully reproduce \nai's extraordinary He\2~1640 line by accounting for enhanced mass-loss from VMS close to the Eddington limit. However, they did not generate a tailored model for this SSC.

\subsection{This work}

 In this paper, we use high-quality UV data that have not been previously analysed in combination with state-of-the-art population synthesis models to further investigate if VMS can account for the extreme broad He\2 emission in \nai. In addition, we explore whether VMS can also account for the previously mentioned observation of CDFS131717. 
 
In \S~\ref{sec:2}, we present \nai, summarise previous findings about this object, present the UV observations of \nai~used in this work, and data reduction. In \S~\ref{sec:3}, we describe the models used in our analysis of A1. In \S~\ref{sec:4}, we show the comparison of the models with the observations of A1. In \S~\ref{sec:5}, we discuss our results for A1 and their consequences for the interpretation of the rest-frame UV spectrum of high-z galaxy, CDFS131717. Finally, in \S~\ref{sec:6}, we provide our summary and conclusions.

\onlyinsubfile{
\bibliographystyle{mnras}
\bibliography{REFERENCES/biblio}
\section{Observations and data reduction}\label{sec:2}
\section{Population Synthesis Models}\label{sec:3}
\section{COS observations versus CB19 models}\label{sec:4}
\section{Discussion}\label{sec:5}
\section{Summary and conclusions}\label{sec:6}
}

\section{Target, observations and data reduction}\label{sec:2}

\subsection{\nai}

Nearby starburst galaxy NGC~3125 (Tol 3, distance estimates provided in Table~\ref{tab:ngc3125}) is of particular importance for studies of the upper stellar initial mass function (IMF; e.g., W14) and galaxies at cosmic noon \citep[e.g.][]{2004ApJ...604..153C}. The galaxy has an ionised-gas oxygen abundance of 12+log(O/H)=8.3 \citep{2006MNRAS.368.1822H}, which is consistent with abundances measured in the Large Magellanic Cloud (LMC) via collisionally excited lines \citep{1990ApJS...74...93R, 2017MNRAS.467.3759T} and characteristic of some star-forming galaxies at cosmic noon \citep{2016ApJ...826..159S, 2017NatAs...1E..52A, 2021ApJ...914...19S, 2021MNRAS.505..903C}.  It is dominated by two emission regions, A and B. Region A can be subdivided into ultraviolet (UV)-bright and highly extinguished components A1 and A2, respectively \citep{2006MNRAS.368.1822H}. Given its young age and high mass (2.2\,Myr and $1.9\times10^5\,M_\odot$, respectively; values from this work; see also W14), A1 is considered to be a super star cluster (SSC, \citealt{2001PhDT.......182J, 2020SSRv..216...69A}). Table~\ref{tab:ngc3125} lists main properties of the galaxy, region A, and SSC A1.

 In W14, we present {\it Hubble Space Telescope} (\hst) UV spectra of A1 obtained by PI Leitherer with two spectrographs: i) the Space Telescope Imaging Spectrograph (STIS), using the 2"x50" long slit (effectively a 2"x25" slit due to the size of the STIS FUV MAMA detector) and the G140L grating ($R\sim1000$, i.e., $\sim300$ km\,s$^{-1}$ at 1640\,\AA); and ii) the Cosmic Origins Spectrograph (COS), using the 2.5"\,-\,wide circular Primary Science Aperture (PSA) and the G130M grating ($R\sim20,000$, i.e., $\sim15$ km\,s$^{-1}$ at 1640\,\AA). We also show that the spectroscopically-derived mass of A1 is of the same order of magnitude as the binary-corrected virial mass of star-forming region NGC~2070 in the LMC ($4.5\times10^5\,$M$_\odot$) and  consistent with a mass estimate from photometry \citep{2009AJ....137.3437B}. The STIS observation analysed in W14 shows that A1 is characterised by the presence of a broad He\2~$\lambda$1640 emission line with a large equivalent width ($FWHM=1131\pm40\,$km\,s$^{-1}$ and $EW=4.6\pm0.5\,$\AA, respectively, values from this work) and stellar O V $\lambda$1371 absorption. W14 show that the He\2~emission in the STIS spectrum of A1 cannot be reproduced with standard Starburst99 models \citep{2010ApJS..189..309L}.

\begin{table}
    \centering
    \begin{tabular}{l|l|l|c}
    \hline
    Property & Units & Value & Reference \\
    \hline
         RA$_{\rm N3125}$ & h:m:s & 10:06:33.372 & 1\\ 
         Dec$_{\rm N3125}$ & d:m:s & -29:-56:-5.500 & 1 \\ 
         RA$_{\rm A1}$ & h:m:s & 10:06:33.280 & 2 \\
         Dec$_{\rm A1}$ & d:m:s & -29:-56:-6.800 & 2 \\
         Redshift & - & 0.003712 & 3 \\
         Distance 1 & Mpc & 11.5 & 4a \\
         Distance 2 & Mpc & $14.84\pm1.04$ & 4b\\
         $E(B-V)_{\rm{MW}}$ & mag & 0.073 & 5 \\
         $E(B-V)_{\rm{A1}}$ & mag & 0.130 & 6 \\
         12+log(O/H)$_{\rm A}$ & dex & $8.32\pm0.03$ & 7 \\
         \hline
    \end{tabular}
    \caption{Main properties of NGC~3125, region A, and SSC A1. Reference:\\
    (1) J2000 right ascension (h:m:s) or declination (d:m:s) of galaxy from Strasbourg astronomical Data Center (\url{http://cdsweb.u-strasbg.fr/}).\\
    (2) J2000 right ascension (h:m:s) or declination (d:m:s) of A1 used for COS observations from \url{https://archive.stsci.edu/missions-and-data/hst}.\\
    (3) Redshift from NASA/IPAC Extragalactic Database (NED, \url{https://ned.ipac.caltech.edu/}).\\
    (4a) Value derived from the Galactic Standard of Rest velocity using a Hubble constant of $H_0 = 75$\,km\,s$^{-1}$\,Mpc$^{-1}$, obtained by \citet{1999A&A...341..399S}, and adopted in W14 for comparison of the SSC masses and numbers of O and WR stars with previous works.\\
    (4b) Virgo + GA + Shapley distance \citep{2000ApJ...529..786M} from NED for cosmological parameters: $\Omega_{\rm{matter}}$ = 0.308, $\Omega_{\rm vacuum}$ = 0.692, and $H_0$ = 67.8 km\,s$^{-1}$\,Mpc$^{-1}$ \\
    (5) Colour excess due to dust in the Milky Way from: \url{https://noirlab.edu}.\\
    (6) Colour excess due to intrinsic dust from W14.\\
    (7) Oxygen abundance in the ionised gas from \citet{2006MNRAS.368.1822H}.    
    }\label{tab:ngc3125}
\end{table}

\onlyinsubfile{
\bibliographystyle{mnras}
\bibliography{REFERENCES/biblio}
}

\onlyinsubfile{
\bibliographystyle{mnras}
\bibliography{REFERENCES/biblio}
}

\subsection{Observations}\label{sec:2.1}

Observations of \nai~were obtained with COS G130M and G160M gratings as part of \hst~programs 12172 (PI Leitherer, on November 2011; lifetime position, LP1) and 15828 (PI Wofford, on October 2020; LP4), respectively. In particular, we use the data sets that are listed in column 1 of Table~\ref{tab:cos_log}. The table gives the  grating, central wavelength, and exposure time of the observations. The spectra were obtained with the PSA. To provide continuous coverage from $\sim1150-1800$\,\AA, we used two central wavelengths (cenwave)  with each grating. To improve the quality of the spectra, we used  four focal-plane positions with each cenwave. The observations have a wavelength-dependent resolving power in the range, $R\sim15,000-20,000$.

\begin{table}
    \centering
    \begin{tabular}{l | l | c | l | l}
        \hline
        Dataset & Grating & Cen Wave / \AA & Exp Time / s & PID \\
        \hline
         LBJS01010 & G130M & 1300 & 2549.184 & 12172 \\
         LBJS01020 & G130M & 1318 & 2971.200 & 12172 \\
         LE4GP2010 & G160M & 1600 & 4880.640 & 15828 \\
         LE4GP2020 & G160M & 1623 & 5188.640 & 15828 \\
        \hline
    \end{tabular}
    \caption{Summary of COS spectroscopic observations.}
    \label{tab:cos_log}
\end{table}

\onlyinsubfile{
\bibliographystyle{mnras}
\bibliography{REFERENCES/biblio}
}

\begin{figure}
    \centering
    \includegraphics[width=0.49\textwidth]
    {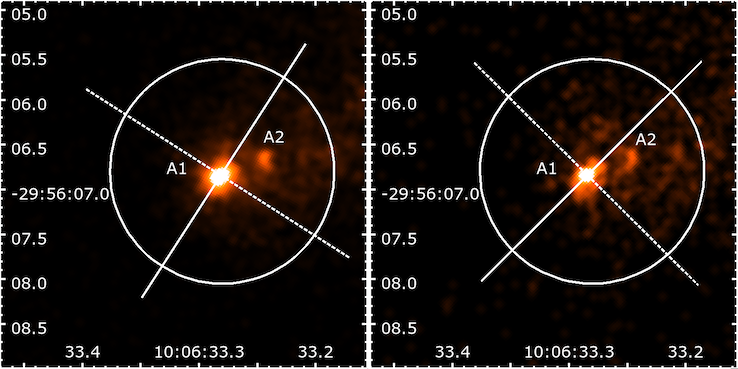}
    \caption{COS NUV MIRRORA TA images of programs 12172 (left; dataset: LBJS01IOQ) and 15828 (right; dataset: LE4GP2ZCQ), with COS footprint overlaid (circle). North is up and east is to the left. The dashed lines indicate the PA angles of the programs, (149 and 135 degrees, respectively). The exposure times are 30 and 6 s, respectively.  }
    \label{fig:ta_images}
\end{figure}

As described in \citet{2022cosi.book...14J}, offsets in flux between different gratings and cenwaves can be observed in COS observations of extended objects obtained with different
position angles (PAs). We inspected the individual exposures of both gratings in the overlapping regions and found no flux offsets between the different sets. This
is in agreement with the fact that both sets were observed at comparable PAs (135 and 149 deg for G130M and G160M, respectively). The target was acquired using MIRROR A. The near-UV (NUV) acquisition images from programs 12172 (G130M) and 15828 (G160M) are shown in the left and right panels of Fig.~\ref{fig:ta_images}, respectively. In spite of the lower exposure time used in the TA corresponding to program 15828, the fainter A2 cluster is clearly seen in the right panel of the figure. According to \cite{2006MNRAS.368.1822H} the total colour excesses of A1 and A2 are: $E(B-V) \sim0.24$ and $\sim0.58$ mag, respectively; and A1 has an NUV flux in the \hst~FOC F220W band that is approximately 3.2 times larger than that of A2.

\onlyinsubfile{
\bibliographystyle{mnras}
\bibliography{REFERENCES/biblio}
}

\subsection{Data Reduction}\label{sec:2.2}

The observations were retrieved from the Mikulski Archive for Space Telescopes (MAST) and calibrated on-the-fly using \hst's~CALCOS code version V3.4.0 \citep{2021cosd.book....5S}. The latest version of the CALCOS pipeline does not provide co-added products combining different gratings and central wavelengths. We further process the individual COS x1d files using the IDL code developed by the COS Guaranteed Time Observer Team \citep{2010ApJ...720..976D}. The individual x1d spectra were weight-combined (using the exposure times as weighting factors) by interpolating onto a common wavelength vector. Lastly, we bin the combined spectrum by a single COS resolution element (1 resel = 6 pixels), corresponding to the nominal point-spread function. We checked the wavelength calibration accuracy by comparing the position of the MW absorption line Si\2\,$\lambda$1190.42 \AA~and Al\2\,$\lambda$1670.79. We find offsets of about $\sim0.05$ \AA~that are within the wavelength accuracy of COS (0.06 \AA).

\begin{figure*}
    \centering
    \includegraphics[width=\textwidth]{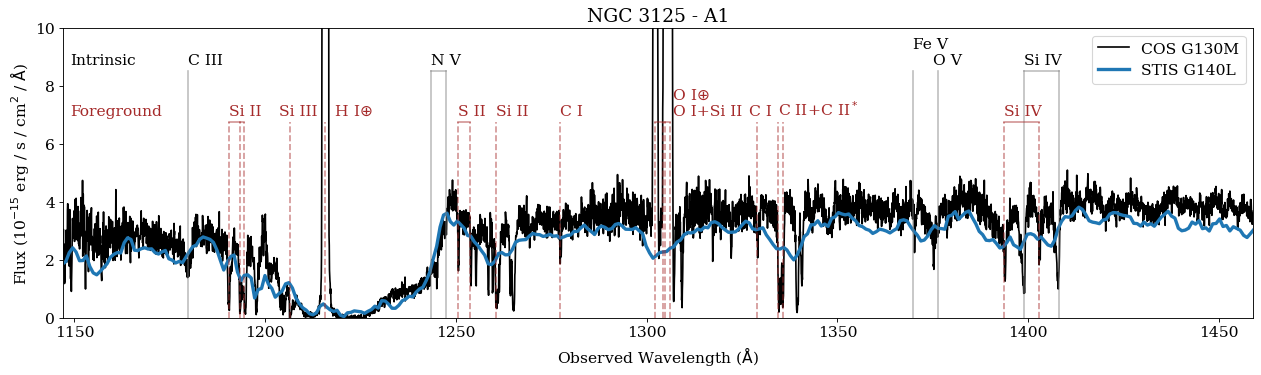}\\
    \includegraphics[width=\textwidth]{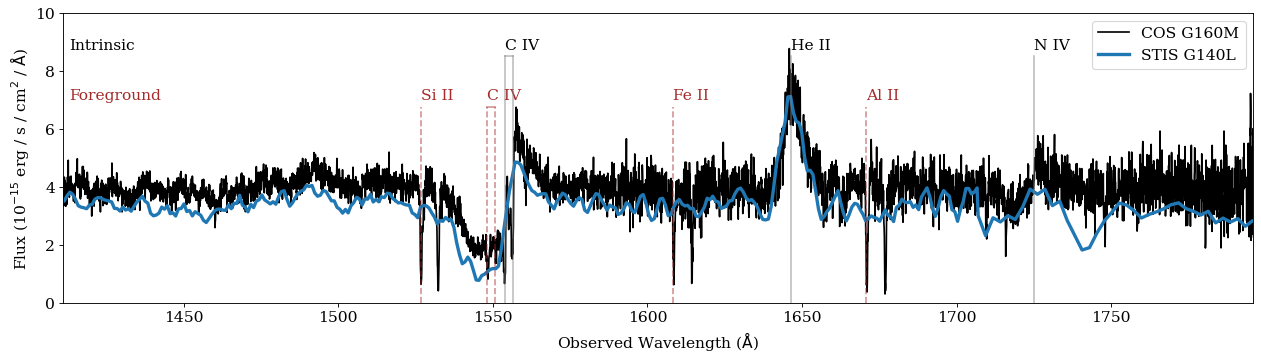}
    \caption{\nai˜spectra, uncorrected for any reddening due to dust and redshift, obtained with COS G130M + G160M (black curve) and  STIS G140L  (blue curve), using the 2.5" circular and 2"x25" long-slit apertures, respectively. We mark the positions of intrinsic stellar-wind and photospheric features with solid vertical grey lines. The cores of some of these features are contaminated with ISM absorptions. The ions that the features originate from are indicated with black characters. We also mark the positions of strong foreground contaminating features with dashed vertical brown lines, including the geocoronal H\1~and~O\1~emission lines, which are indentified by the Earth's symbol. Note the broad absorption at $\sim1216\,$\AA~due to the H\1~\LYA~ absorptions of the MW and the galaxy. The limits of the top and bottom panels are [1147, 1459] and [1411, 1796] \AA, respectively, and correspond to the limits covered by the G130M and G160M observations.}
    \label{fig:cos_vs_stis}
\end{figure*}

Figure~10 of W14 shows a comparison between COS G130M and STIS G140L spectra of \nai. Figure~\ref{fig:cos_vs_stis} shows a similar figure where the wavelength range has been extended in order to include the unpublished COS G160M observations. In Fig.~\ref{fig:cos_vs_stis}, we identify the strongest foreground and intrinsic spectral features. For this purpose, we use the line list in Table~1 of \cite{2011AJ....141...37L}. The difference in flux between the STIS and COS observation is due to the difference in apertures.

 Figure~\ref{fig:cos_vs_stis} shows that the \nai~profiles of  N\5~$\lambda\lambda1238,1241$ (N\5~1240), C\4~$\lambda\lambda$1548, 1551 (C\4 1550), and He\2~1640; and Milky Way (hereafter MW) + \nai~ H\1~\LYA~absorptions are remarkably similar in the COS and STIS observations. However, the STIS observation has a lower spectral resolution. This makes it hard to see weak narrow ISM absorption lines and to distinguish between intrinsic and foreground ISM lines, except when these are sufficiently strong, as is the case for the C\2, C\2$^*$\,$\lambda\lambda$1334, 1336 and Si\4~$\lambda\lambda$1393, 1403 (Si\4 1400) doublets. Note that the weaker N\4$\lambda$1718 P-Cygni like profile is much clearer in the COS data.

There is a striking similarity between the spectrum of \nai˜that we show in Fig.~\ref{fig:cos_vs_stis} and the spectrum of  VMS, R136a3, which is shown in fig. 5 of \citet{2010MNRAS.408..731C} (top panel). The VMS spectrum was obtained with the Goddard High Resolution Spectrograph (GHRS, an instrument that is no longer on board of \hst). For reference, R136a3 is located in the LMC, is classified as a hydrogen-rich WN5h star, and has an initial mass of $155^{+25}_{-18}$ \citep{2022ApJ...935..162K}. 

\onlyinsubfile{
\bibliographystyle{mnras}
\bibliography{REFERENCES/biblio}
}

\subsection{Normalisation and Ly$\alpha$ correction}\label{sec:4.1}

Removing the \LYA~absorption profiles of the Milky Way and NGC~3125 from the COS observations is a critical step for finding an accurate model fit to the observations. This is because these absorptions severely contaminate the N\5~1240 stellar doublet. We correct for \LYA~absorption by adopting a similar approach and software as those described in \citet{2020ApJ...892...19H, 2021ApJ...908..226H} and \citet{ 2022AJ....164..208S}.

As a first step, we normalise the COS spectrum by interpolating between nodes manually positioned to avoid stellar and ISM features. To guide the normalisation and location of the nodes, we use a \texttt{Staburst99} \citep{1999ApJS..123....3L, 2014ApJS..212...14L} instantaneous burst model of LMC metallicity. We adjust the tension of the cubic spline between the nodes as needed to closely match the rectified luminosities of the model.

Using the normalised spectrum we fit Voigt profiles to the Lyman $\alpha$ absorption using the Python software \texttt{VoigtFit} v.3.13.9 \citep{2018arXiv180301187K}. We note that given the radial velocity of NGC~3125, the \LYA~absorption originating from the MW is blended with the galaxy absorption. We fit both \LYA~absorption components (MW and NGC~3125) simultaneously. We set the b value for both MW and NGC~3125 to 0, since the line is in the damped part of the curve of growth; and the velocity for the MW is set to 0 km/s.  The \texttt{VoigtFit} software allows users to incorporate the line-spread function (LSF) information to account for the wavelength-dependent resolution intrinsic to the instrument. Given that \nai~is observed to be an extended source, we convolve the COS LSF profiles with the $FWHM$ of the target in the dispersion direction, as measured from the COS acquisition images, and use the broadened LSFs to specify the resolution of the observations within \texttt{VoigtFit}. The final spectral resolution of the observations is derived using equation (1) in \citet{2020ApJ...892...19H}. We obtain the best \LYA~fit when masking contaminating stellar and ISM absorption features, particularly the broad N\5~1240  absorption on the red wing of \LYA. We fit for the hydrogen column densities NH\1(MW), NH\1(NGC~3125), and the velocity of NGC~3125.  We measured a column density of H\1 of log NH\1 = $21.752\pm0.126$ cm$^{-2}$ for NGC~3125. The inferred NH\1(MW) agrees with the measured 21 cm value from the all-sky map from HI4PI (\url{http://cdsarc.u-strasbg.fr/viz-bin/qcat?J/A+A/594/A116}). We divide the normalised COS spectrum by the best \LYA~fit to remove the contaminating profile.

\begin{figure*}
    \centering
    \includegraphics[width=\textwidth]{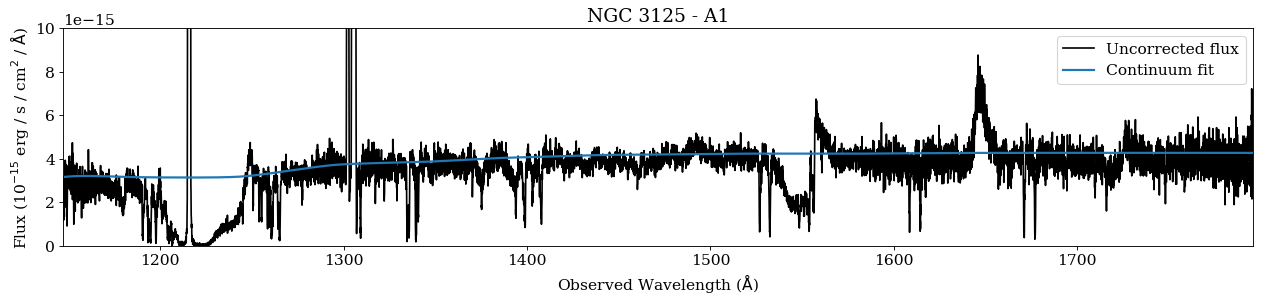}\\
    \includegraphics[width=\textwidth]{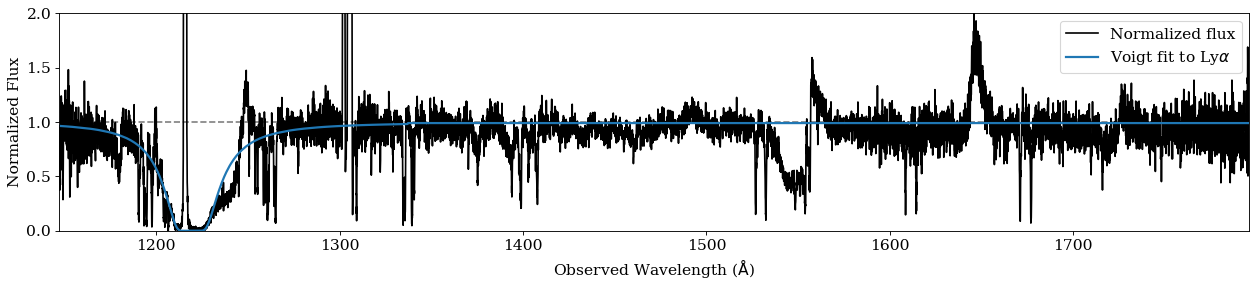}\\
    \includegraphics[width=\textwidth]{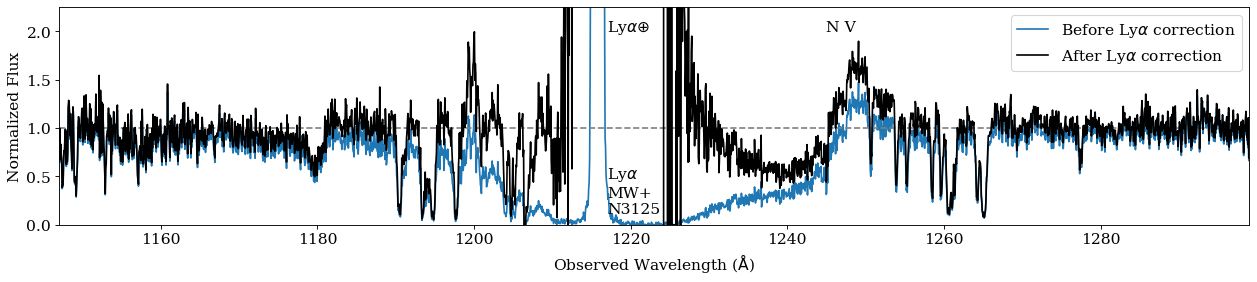}\\
    \caption{Plots corresponding to steps we follow to normalise and correct for \LYA~the COS G130M + G160M spectrum. Top panel.-- COS spectrum (black curve) and fit to the continuum (blue curve). Middle panel.-- Normalised COS spectrum resulting from dividing the flux by the fit to the continuum. Bottom panel.-- Comparison of normalised spectra before and after correcting for the \LYA~absorptions (blue and black curves, respectively).} 
    \label{fig:normalisation_lya}
\end{figure*}

In Fig.~\ref{fig:normalisation_lya}, we show the steps followed to normalise and correct the COS G130M + G160M spectrum for the \LYA~absorptions. The top panel shows the fit to the observed continuum. The middle panel shows the normalised  spectrum that results from dividing the observed spectrum by the fit to the continuum. It also shows the fit to the \LYA~absorption components. Finally, the bottom panel shows the comparison between the normalised spectra before and after correcting for the \LYA~absorptions. The bottom panel clearly shows that the strength of the emission component of the P-Cygni like profile of N\5~$\lambda\lambda$1238, 1242 becomes stronger after the \LYA~correction.

\onlyinsubfile{
\bibliographystyle{mnras}
\bibliography{REFERENCES/biblio}
}

\section{Population Synthesis Models}\label{sec:3}

In W14, we speculated that VMS are present in \nai. However, it was not until recently that population synthesis models accounting for VMS became available \citep[e.g., ][]{2021MNRAS.503.6112S, 2022A&A...659A.163M}.  \cite{2022A&A...659A.163M} were the first to reproduce A1's He\2 line with VMS approaching the electron-scattering Eddington limit. In their modelling, they use a BPASS SSP with single stars corresponding to $Z=0.006$, age = 2 Myr, and upper mass limit of the IMF of  100 M$_\odot$. They supplement this SSP with VMS models generated with stellar evolution tracks from STAREVOL \citep{2000A&A...358..593S,2012A&A...543A.108L,2019A&A...631A..77A}, stellar atmospheres from CMFGEN \citep{1998ApJ...496..407H}, and the mass loss rates of \cite{2021A&A...647A..13G}. Although their solution overproduces emission in N\5~1240 (see their figure 11), this could be because they do not correct the STIS observation for the nearby LyA absorptions of the MW and NGC~3125. The C\4~1550 emission component is also overproduced in their model. Nevertheless, the fact that they are able to reproduce A1's extreme He\2 emission constitutes a great accomplishment.

The COS observations that we use in this work have higher spectral resolution and SNR in the continuum compared to the STIS observation used by \cite{2022A&A...659A.163M}. They also cover a longer wavelength range, and in particular, the He\2~1640 line at higher spectral resolution. We use these higher quality observations to test the Charlot \& Bruzual (hereafter C\&B) population synthesis models \citep{2019MNRAS.490..978P}. The stellar ingredients used in the C\&B models are explained in detail in \citet[][appendix A, tables 8 to 12]{2022ApJS..262...36S}. These models cover in detail the evolution of O and B stars, including the Wolf-Rayet (WR) phase, and account for the contribution of VMS of up to  $300\,M_\odot$.
The models use the PARSEC stellar evolution tracks of \cite{2015MNRAS.452.1068C}, which are for single, non-rotating stars. The tracks take into consideration the formulation of mass loss rates by \cite{2011A&A...531A.132V}. In summary, for atmospheres and synthetic spectra of massive stars on the Main Sequence (MS) that are not flagged as WR stars according to the criteria provided in section 3.2 of \cite{2015MNRAS.452.1068C}, WM-Basic \citep{2001A&A...375..161P, 2010ApJS..189..309L} models are used. To complete the wavelength-range coverage, the latter models are supplemented with TLUSTY \citep{2003ApJS..146..417L,2007ApJS..169...83L} models. If the star is flagged as WR according to the previous criteria, then C\&B use the model atmospheres and synthetic spectra of WR stars from the Potsdam group (PoWR, Gr\"{a}fener, Hamann, Sander, Shenar, Hainich, Todt et al.),  regardless of whether the star is on the MS or elsewhere. A description of how PARSEC and PoWR models are coupled is specified in the appendix of \cite{2019MNRAS.490..978P}. Note that the mass loss rates of WR stars from the PARSEC tracks are higher than the values predicted by \cite{2011A&A...531A.132V} for stars of the same mass. However, for MS O stars they are consistent. Inhomogeneities in the wind, i.e., clumping in the wind is not accounted for. Since the observation is dominated by cluster A1, we assume a Simple Stellar Population (SSP), i.e., a single burst of star formation. In addition, we exclude the contribution of the ionized gas, as no nebular emission lines are present in the spectra. We leave the metallicity and age of the stellar population as free parameters. We do not include the reddening due to dust in the model because the comparison with the observation is done in normalised units. We try models of two metallicities that approach that of the ionised gas of NGC~3125, Z=0.006 and Z=0.008. For this particular application, we sample the stellar spectra in the C\&B models in the UV range (5 to 3541.4 \AA) every 0.1 \AA\ using the {\it SpectRes} tool provided by \citet{2017arXiv170505165C}. The total number of wavelength points in each resampled C\&B spectrum is then 46,226, instead of the 16,902 points in the standard distribution \citep[][table 12]{2022ApJS..262...36S}. 

In Fig. \ref{fig:cb19_1640}, we show corresponding predictions for the profile shape of the He\2 1640 line.  We show results for the above two metallicities and cases where the upper IMF mass limit, $M_{\rm up}$, is equal to 100 and 300 $M_\odot$ (hereafter, M100 and M300 models, respectively). We only show ages ranging from 1 to 7 Myr, as for SSPs with single stars, the He\2˜line is expected to have an emission component at ages of $<5$ Myr, when VMS and WR stars are present. The curves show the normalised luminosity. In each panel, we plot models of different ages given in Myr by the legend on the left. Each row of the figure shows predictions for different age ranges. The legends on the right give the maximum luminosity reached within each panel and its corresponding age. 

Figure \ref{fig:cb19_1640} shows that at the two metallicities, the M300 models yield larger amplitudes of the He\2˜line profiles, relative to the M100 models. This occurs between 2 and 3.0 Myr. In particular, the largest amplitude is reached for the combination of M300, $Z=0.008$, and 2.2 Myr. Note that the WM-Basic models dominate the P-Cygni like He\2 profiles when non-WR MS stars are present in the population but that the PoWR models dominate the He\2 emission when WR stars are present. This is illustrated in appendix \S~\ref{appendix}.

\begin{figure*}
    \centering
    \includegraphics[width=0.5\textwidth]{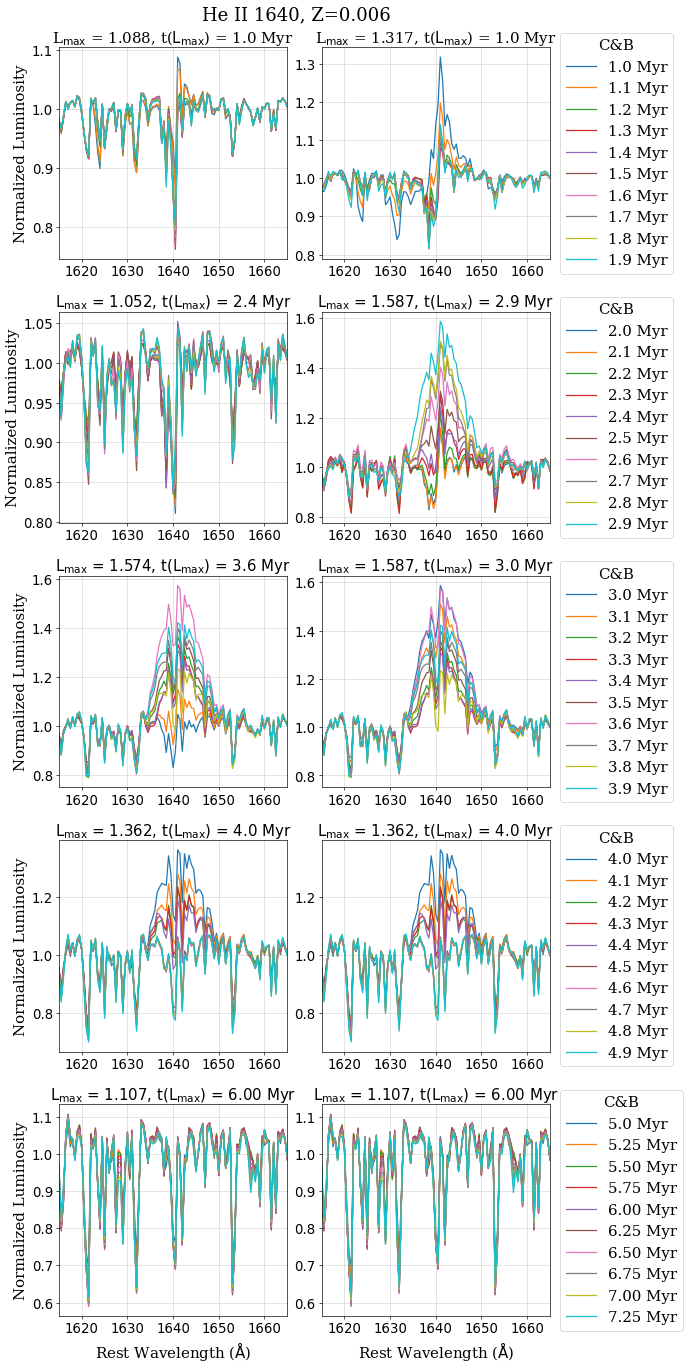}%
    \includegraphics[width=0.5\textwidth]{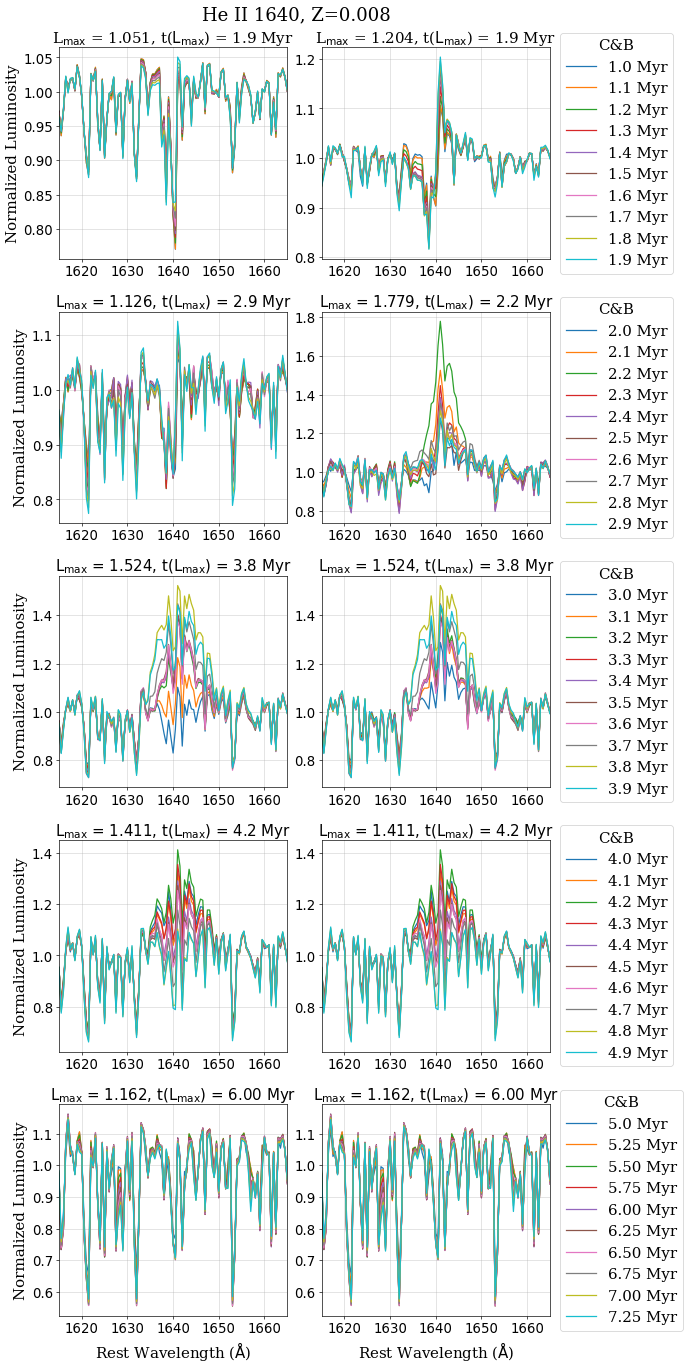}
    \caption{First two columns--. C\&B He\2$\,1640$ line-profile predictions for SSPs with $Z=0.006$ and $M_{\rm up}/M_\odot=100$ {\bf (1st column)} or $M_{\rm up}/M_\odot=300$ {\bf (2nd column)}. We use a linear fit between continuum windows [1615,\,1620] and [1660,\,1665] to normalise the He\2 lines. In each panel, the curves of different colours correspond to models of 10 different ages (except for the last row, which has 9), as given by the right-hand side legends. We give in each panel, the maximum luminosity in the wavelength range [1620,\,1660] $\AA$ that is reached in the panel, i.e., the amplitude of He\2 line when it is in emission; and the corresponding age. Models are available for ages in logarithmic space, which is why in the last row the time step changes and there are less curves. Last two columns--. Similar but for $Z=0.008$. The maximum amplitude of He\2 is obtained for $Z=0.008$, $M_{\rm up}/M_\odot=300$, and 2.2 Myr.}
    \label{fig:cb19_1640}
\end{figure*}

\onlyinsubfile{
\bibliographystyle{mnras}
\bibliography{REFERENCES/biblio}
}

\section{COS observations versus C\&B\ models}\label{sec:4}

\subsection{Nebular contribution check}\label{sec:4.2}

The \hst~ACS/HRC F658N image that is displayed in fig. 1 of  W14 shows that some nebular H$\alpha$ emission is expected within the COS aperture. Since the models we use do not include the contribution of the ionised gas, we use the COS G160M observation to check for the presence of any nebular contribution to the He\2~1640 emission line.    

\begin{figure*}
    \centering
    \includegraphics[width=\textwidth]{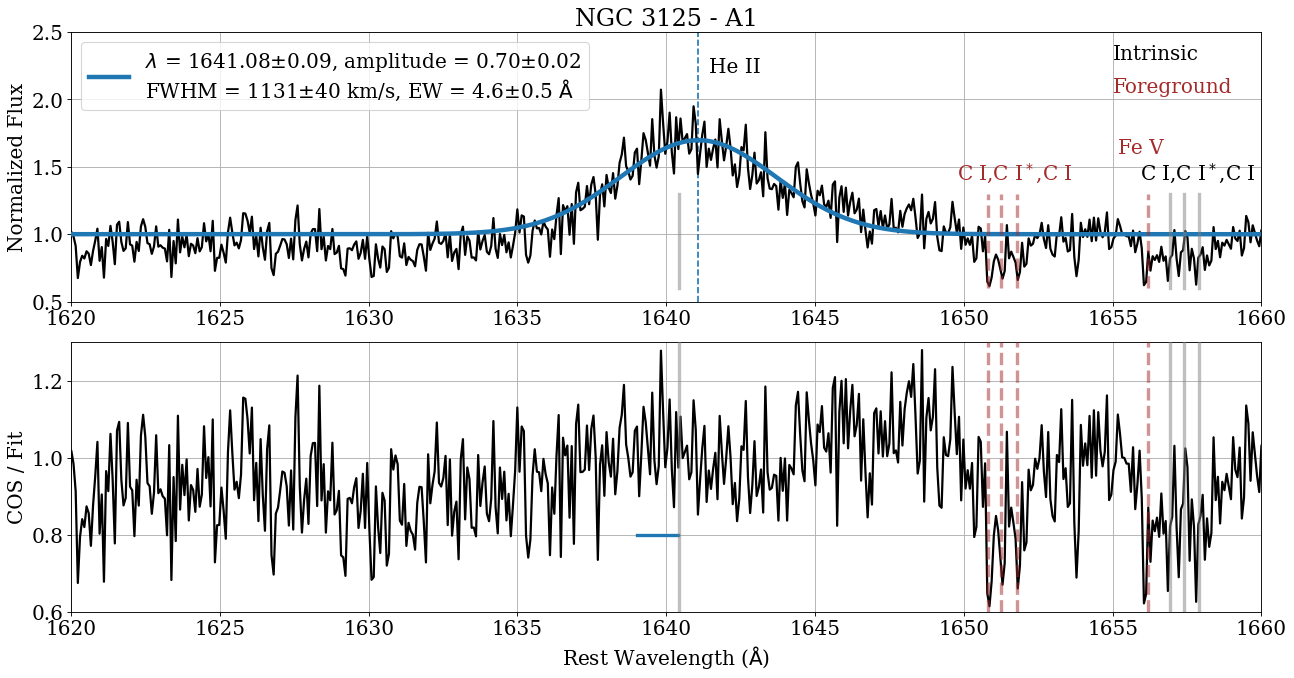}
    \caption{Top panel.--Normalised portion of the spectrum around the intrinsic He\2~$\lambda$1640 line (black curve), and Gaussian fit to the line (blue curve). The spectrum is corrected for the redshift of the galaxy. After this correction, the observed He\2~line, whose position at 1641.08\AA~is marked by a vertical blue-dashed line, shows an additional redshift of $121\pm17\,$km\,s$^{-1}$ relative to the expected rest-frame position at 1640.42 \AA, which is marked by the short vertical grey line. We mark the position of the He\2 line  \AA~with a long, vertical, blue-dashed line {\bf and the rest-frame He\2 wavelength with a short vertical grey line.} The fit parameters are given in the legend. Bottom panel.-- Observation divided by fit (black curve). In both panels, we use thick brown-dashed and grey-solid vertical lines to mark the positions of foreground and intrinsic lines. The corresponding ions are indicated in the top panel.}
    \label{fig:he2_fit}
\end{figure*}

The top panel of Fig.~\ref{fig:he2_fit} shows a single-component Gaussian fit (blue curve) to the  normalised COS observation around the He\2~line  (black curve). The best-fit parameters are given in the top-left legend of the panel. The emission line is broad ($FWHM=1131\pm40\,$km\,s$^{-1}$), strong ($EW=4.6\pm0.5$\,\AA), and redshifted by $121\pm17\,$km\,s$^{-1}$ in the rest-frame of the galaxy (we provide the galaxy's redshift in Table~\ref{tab:ngc3125}). 

\begin{figure*}
    \centering
    \includegraphics[width=0.75\textwidth]{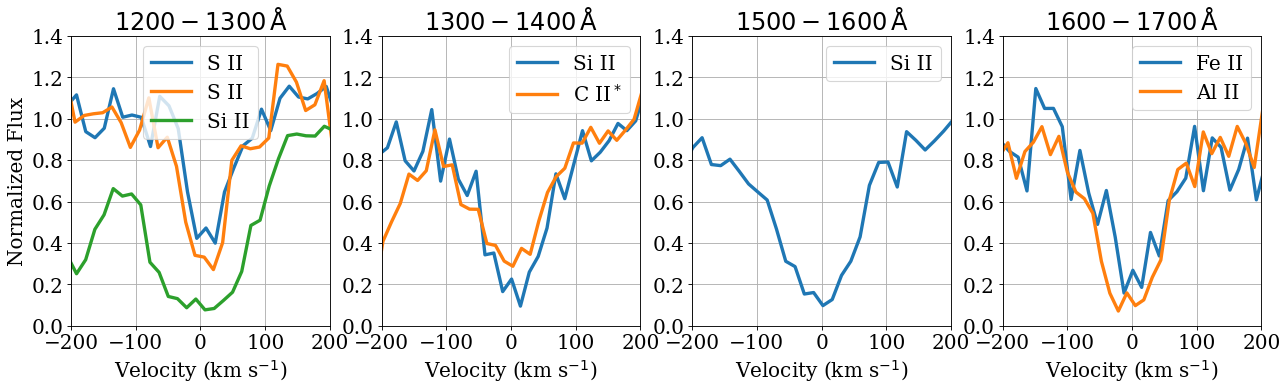}
    \caption{Selection of ISM lines along A1's line of sight that are: free of strong contamination by nearby lines, from the ions given by the legends, and present in the wavelength intervals given by the titles of the panels. According to \citet{2011AJ....141...37L}, the rest-wavelengths of the lines from each ion are: S\2 at 1250.58 and 1253.81 \AA; Si\2 at 1260.42, 1304.37, and 1526.71 \AA; C\2$^*$ at 1335.71 \AA, Fe\2 at 1608.45 \AA; and Al\2 at 1670.79 \AA. We show the redshift-corrected spectra ($z=0.003712$). The centroids of all lines are at rest.}
    \label{fig:redshift}    
\end{figure*}

As shown in Fig.~\ref{fig:redshift}, after correcting the galaxy for redshift, the intrinsic ISM lines are at rest.  The fact that the He\2 line is redshifted relative to these lines is likely because VMS dominate the He\2 emission, as in R136 (e.g., fig. 12 of \citealt{2016MNRAS.458..624C}), and these stars  have P-Cygni like He\2 profiles. In the models with $Z=0.008$ shown in Fig.~\ref{fig:cb19_1640}, the predicted He\2 line profiles of the SSPs are P-Cygni like at ages of $<3\,$Myr. Thus, in the COS observation, we are likely seeing the redshifted emission component of the P-Cygni profile. The corresponding blueshifted absorption is expected to be weak and is buried in the noise.

The bottom panel of Fig.~\ref{fig:he2_fit} shows the observation divided by the Gaussian fit to the He\2 line. There appears to be a decrease in the flux blueward of the He\2 emission peak that could be due to the P-Cygni absorption component. There is also a small flux excess on the blue side of the broad component (region marked by horizontal blue line in the bottom panel) at the level of $\sim10\%$. However, the He\2 profile is clearly dominated by stellar-wind emission. We attribute the lack of significant nebular emission in the COS observation to the fact that there is no significant escape of \hep-ionising photons from the VMS stars in A1, as expected from the fact that due to their high mass loss their winds are dense \citep[e.g.,][]{2021A&A...647A..13G} and absorb these photons. 

The lower value of the He\2 $EW$ ($\sim5$ \AA) relative to that previously obtained from the STIS observation ($\sim7$ \AA, W14) is due to the better fit to the continuum near the He\2~line. In particular, with the higher sensitivity and spectral resolution of COS, the weak foreground and intrinsic C\1~and C\1$^*$ lines are detected and resolved. In the STIS observation, the fit to the continuum goes right through these lines, enhancing the line flux and hence the $EW$ of He\2.

\onlyinsubfile{
\bibliographystyle{mnras}
\bibliography{REFERENCES/biblio}
}

\subsection{Best-fit model}\label{sec:4.3}

In W14, none of the models available at the time could reproduce the strength of the He\2 1640 line, except for a model with an unrealistically flat IMF. Now that we have corrected the N\5 profile for \LYA~absorption,  confirmed that there is no significant nebular contamination, and normalised the spectrum, we proceed with the comparison of the observations with the model that predicts the strongest He\2 line, which as shown in Fig.~\ref{fig:cb19_1640}, is the M300 models with $Z=0.008$ and 2.2 Myr. The comparison of the COS observation and the C\&B models is done after resampling the observation onto the model wavelength grid. We use Python's function "spectres" \citep{2017arXiv170505165C} for this purpose.

Figure~\ref{fig:cos_vs_cb19} shows the comparison between the COS G130M + G160M spectrum and the above model. Note that interstellar absorption lines originating in the Milky Way and NGC~3125-A1, which are not included in the model, have not been removed from the observations. On the other hand, the model predicts a P-Cygni like \LYA~profile in the region of the observation that is contaminated with geocoronal \LYA~emission that fills the COS aperture.

The model is not only able to reproduce the He\2 1640 line, but also provides very reasonable fits to the C\3 $\lambda$1175, N\5 1240, and C\4 1550 stellar-wind features. On the other hand, the optimum fit fails to reproduce both the O\5 1371 line, which is indicative of youth \citep[early O stars; e.g.,][]{1995yCat.3115....0W, 
2016MNRAS.458..624C} or presence of WC stars \citep[e.g.,][]{1998A&A...329..190G, 2002A&A...392..653C}); or Si\4 $\lambda\lambda$ 1394, 1403, which is a key indicator of evolved OB supergiants. The mismatch between the observed O\5 and Si\4 and the best fit is likely due to failures of the theoretical models (e.g., \cite{2004ApJ...606..497G} found discrepancies between WMBasic predictions and UV spectra obtained with the {\it International Ultraviolet Explorer}). Note that metal lines, e.g., O\5 1371 and N\4 $\lambda 1718$ are sensitive to clumping in the stellar wind \citep{2005A&A...438..301B}.  Since the C\&B models are based on stellar atmosphere models without wind inhomogeneities, sensitivity to clumping may explain why some of the above features are not as well fitted. Note that for individual stars, the clumping factor is generally used to get a consensus between UV and optical diagnostics of mass loss. If the mass-loss rates are wrong, the line strengths of wind-affected lines will be wrong and so will be inferences from SSP models that use these lines.

\begin{figure*}
    \centering
    \includegraphics[width=\textwidth]{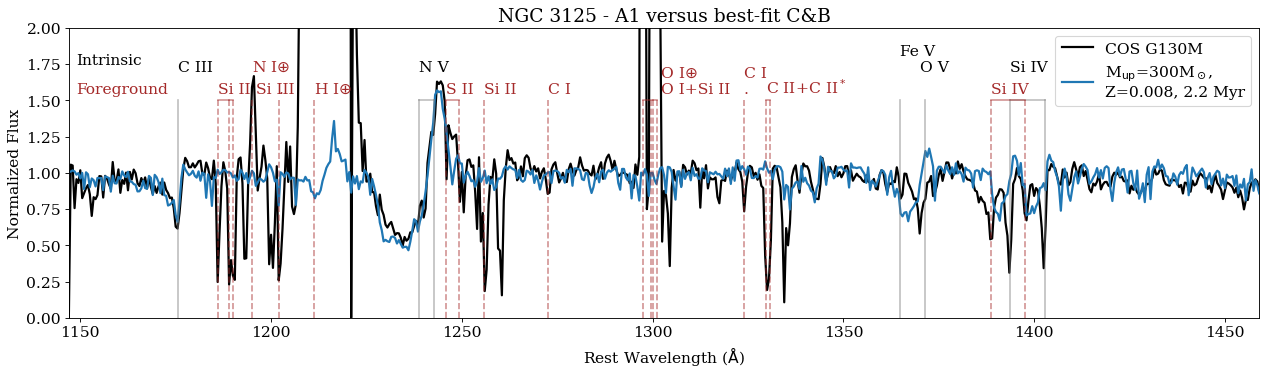}
    \includegraphics[width=\textwidth]{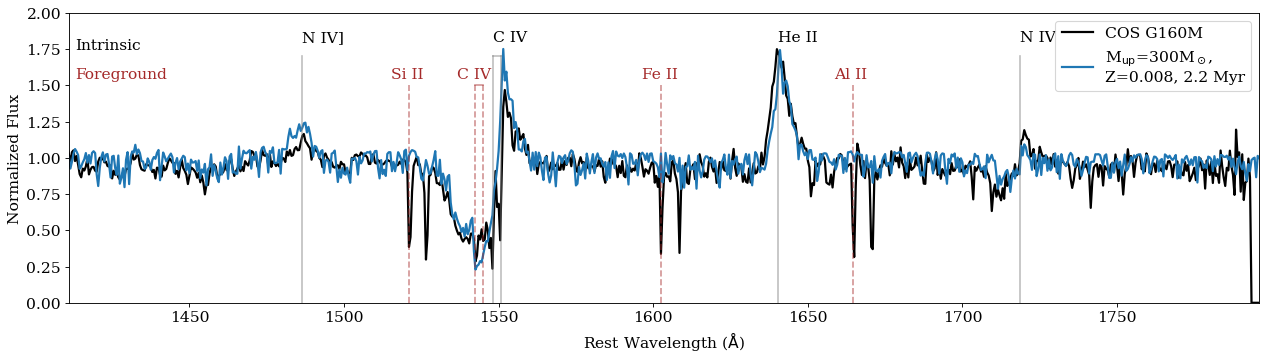}

    \caption{Comparison of CB19 model (blue curve) with COS G130M (top panel) + G160M (bottom panel) observation (black curve). We mark the positions of intrinsic and foreground lines with solid- and dashed-vertical lines, respectively, which we label with black and brown ion designations, respectively. To avoid overcrowding the plot, we do not identify the intrinsic ISM lines which are located to the right of their MW counterparts. We use the Earth´s symbol to identify airglow emission lines.} 
    \label{fig:cos_vs_cb19}
\end{figure*}

\begin{figure*}
    \centering
    \includegraphics[width=\textwidth]{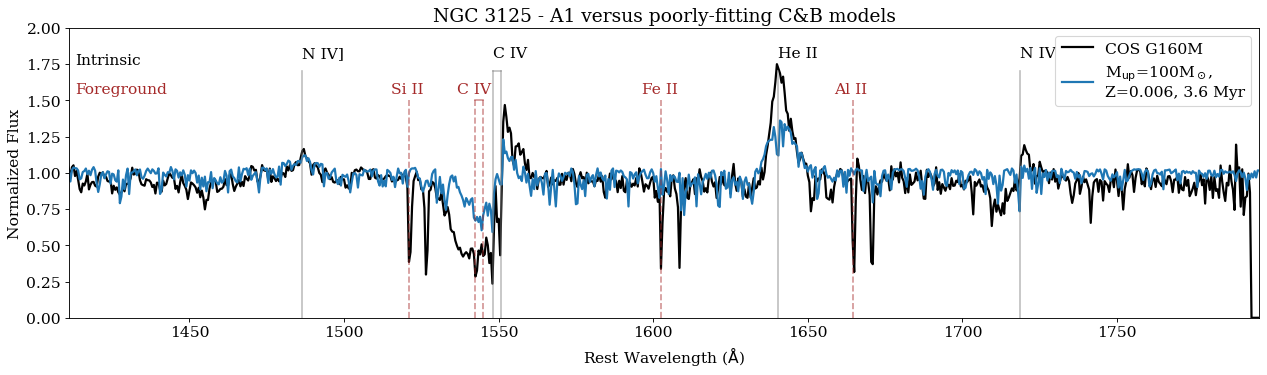}
    \includegraphics[width=\textwidth]{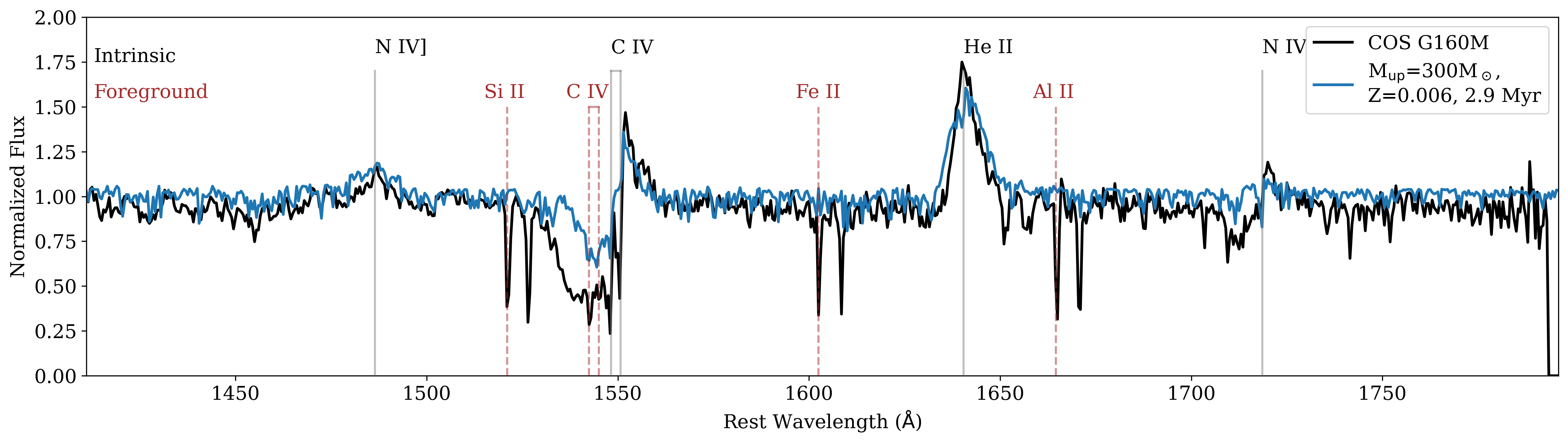}
    \includegraphics[width=\textwidth]{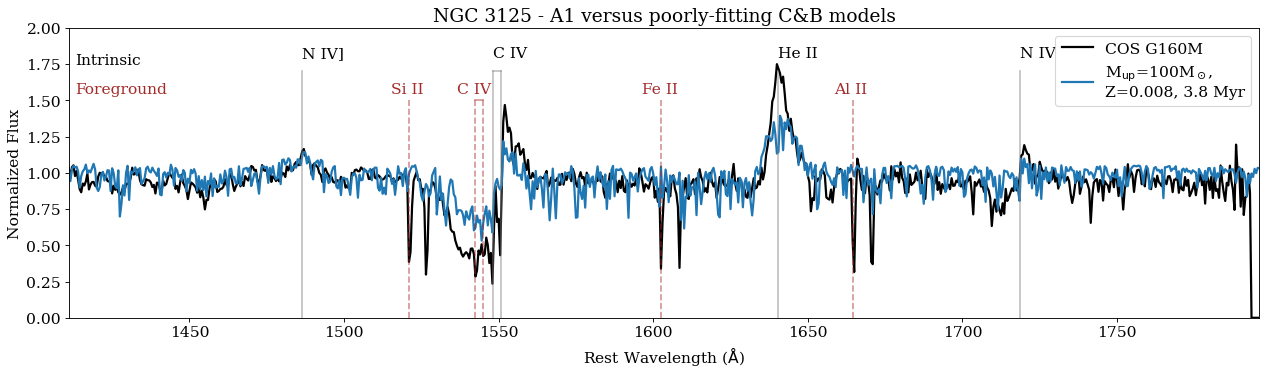}
    \caption{Similar to Fig. \ref{fig:cos_vs_cb19} but for the other combinations of metallicity and $M_{\rm up}$ that are shown in Fig. \ref{fig:cb19_1640}, at the age of maximum amplitude of the He\2 $\lambda$1640 line. We only show the wavelength range covered by G160M. The blue and black curves are the models and observations, respectively. None of the models shown fit simultaneously the C\4 and He\2 features. If we divide the observation by the models and consider the minimum of the stellar C\4 absorption and He\2 emission of the models, it is between $\sim40-50\%$ and $\sim5-30\%$ too weak, respectively, relative to the observation.} 
    \label{fig:cos_vs_other_cb19}
\end{figure*}

For comparison, Fig. \ref{fig:cos_vs_other_cb19} shows models with other combinations of metallicity and $M_{\rm up}$, at the age of maximum amplitude of the He\2 1640 line, which are also shown in Fig. \ref{fig:cb19_1640}. Such models are unable to simultaneously reproduce the stellar C\4 1550 absorption and He\2 1640 emission of \nai. In summary, the M300 model with $Z=0.008$ and 2.2 Myr, is the best fit model.

\onlyinsubfile{

\begin{figure*}
    \centering
    \includegraphics[width=\textwidth]{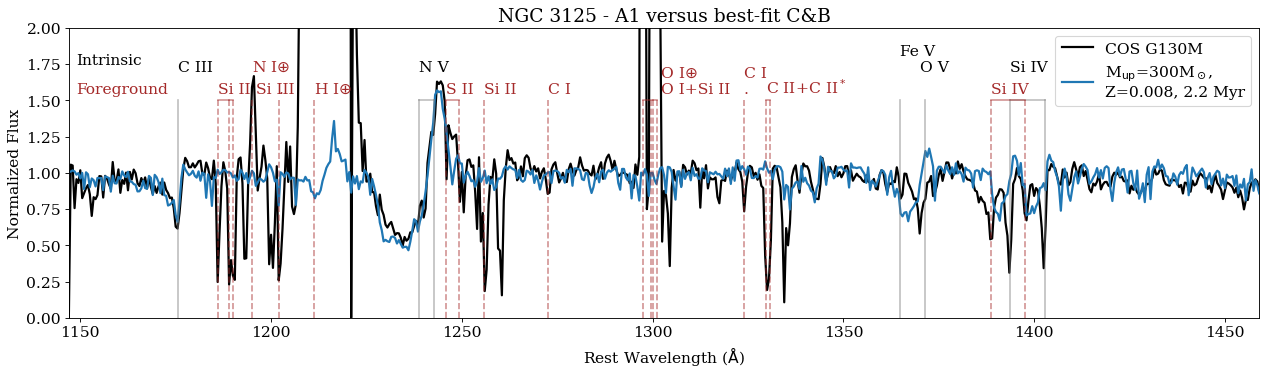}
    \includegraphics[width=\textwidth]{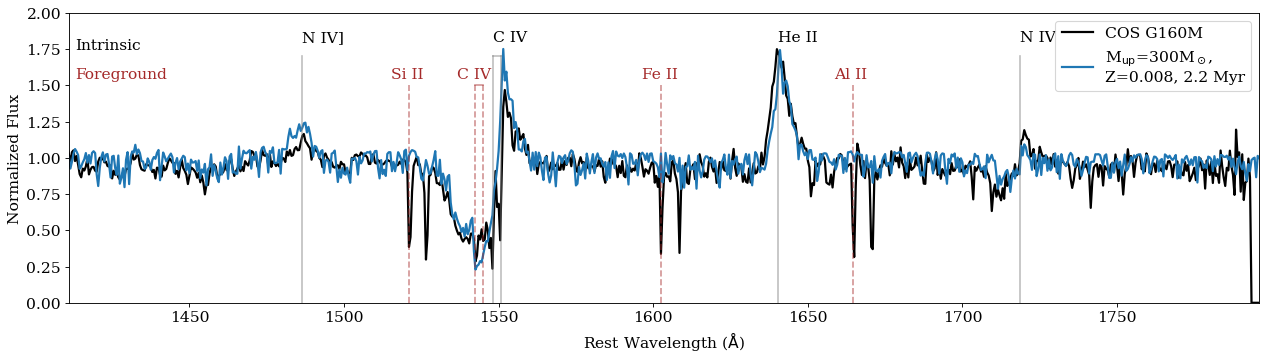}

    \caption{Comparison of CB19 model (blue curve) with COS G130M (top panel) + G160M (bottom panel) observation (black curve). We mark the positions of intrinsic and foreground lines with solid- and dashed-vertical lines, respectively, which we label with black and brown ion designations, respectively. To avoid overcrowding the plot, we do not identify the intrinsic ISM lines which are located to the right of their MW counterparts. We use the Earth´s symbol to identify airglow emission lines.} 
    \label{fig:cos_vs_cb19}
\end{figure*}

}

\onlyinsubfile{

\begin{figure*}
    \centering
    \includegraphics[width=0.5\textwidth]{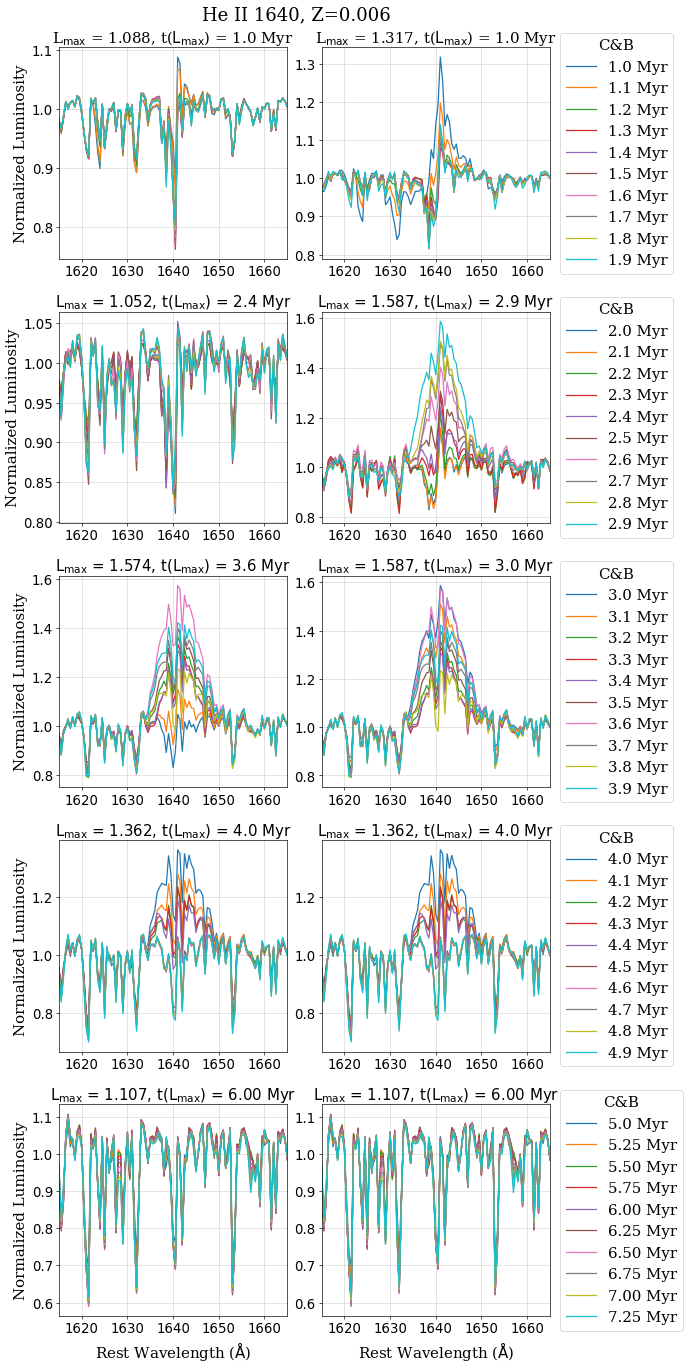}%
    \includegraphics[width=0.5\textwidth]{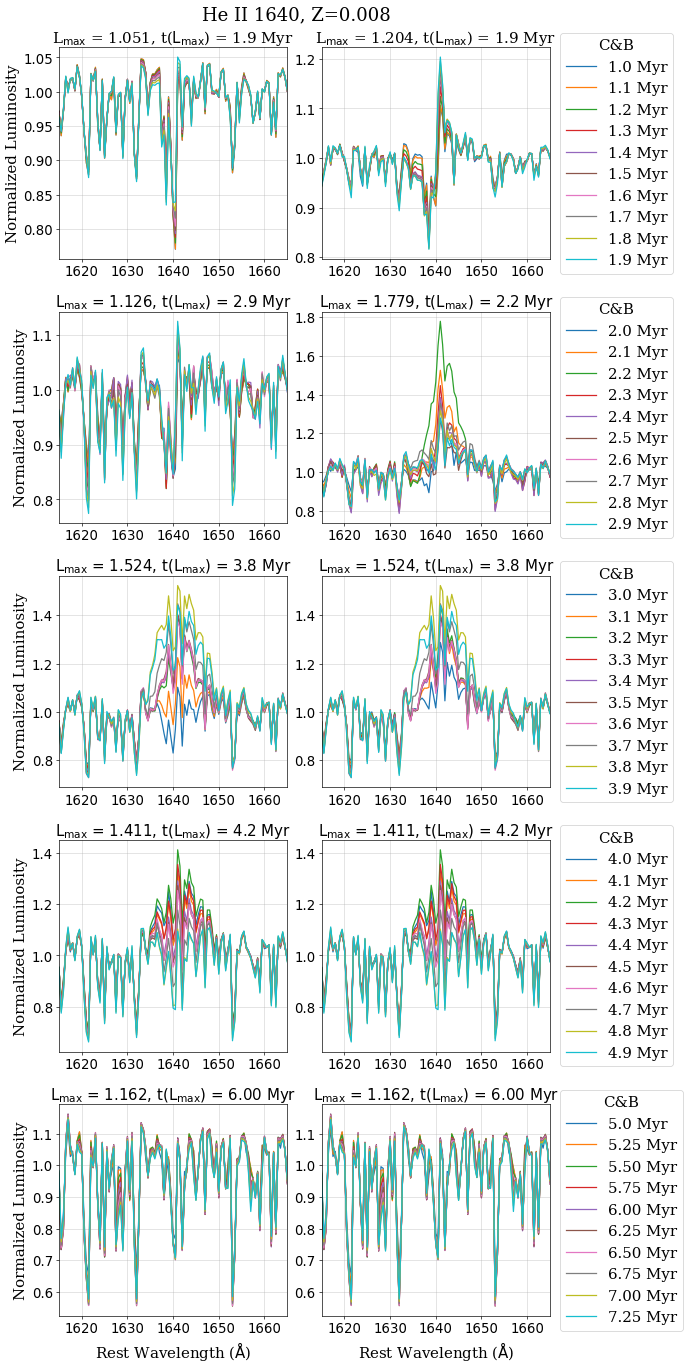}
    \caption{First two columns--. C\&B He\2$\,1640$ line-profile predictions for SSPs with $Z=0.006$ and $M_{\rm up}/M_\odot=100$ {\bf (1st column)} or $M_{\rm up}/M_\odot=300$ {\bf (2nd column)}. We use a linear fit between continuum windows [1615,\,1620] and [1660,\,1665] to normalise the He\2 lines. In each panel, the curves of different colours correspond to models of 10 different ages (except for the last row, which has 9), as given by the right-hand side legends. We give in each panel, the maximum luminosity in the wavelength range [1620,\,1660] $\AA$ that is reached in the panel, i.e., the amplitude of He\2 line when it is in emission; and the corresponding age. Models are available for ages in logarithmic space, which is why in the last row the time step changes and there are less curves. Last two columns--. Similar but for $Z=0.008$. The maximum amplitude of He\2 is obtained for $Z=0.008$, $M_{\rm up}/M_\odot=300$, and 2.2 Myr.}
    \label{fig:cb19_1640}
\end{figure*}

\bibliographystyle{mnras}
\bibliography{REFERENCES/biblio}
}

\section{Discussion}\label{sec:5}

\subsection{Does A1 contain VMS?}\label{sec:5.1}

Several of the stellar-wind features that are observed in the COS spectrum of A1 are also observed in the spectra of classical WR stars. However, the presence of the O\5 $\lambda$1371 absorption line in A1's spectrum suggests that this system might be too young to already have cWR stars. O\5 is common in early O and VMS \citep{2016MNRAS.458..624C}. Although note that a blend of lines including O\5 is used to fit the spectra of in WC stars in \cite{2022ApJ...924...44A}. 

\begin{figure}
    \centering
    \includegraphics[width=0.5\textwidth]{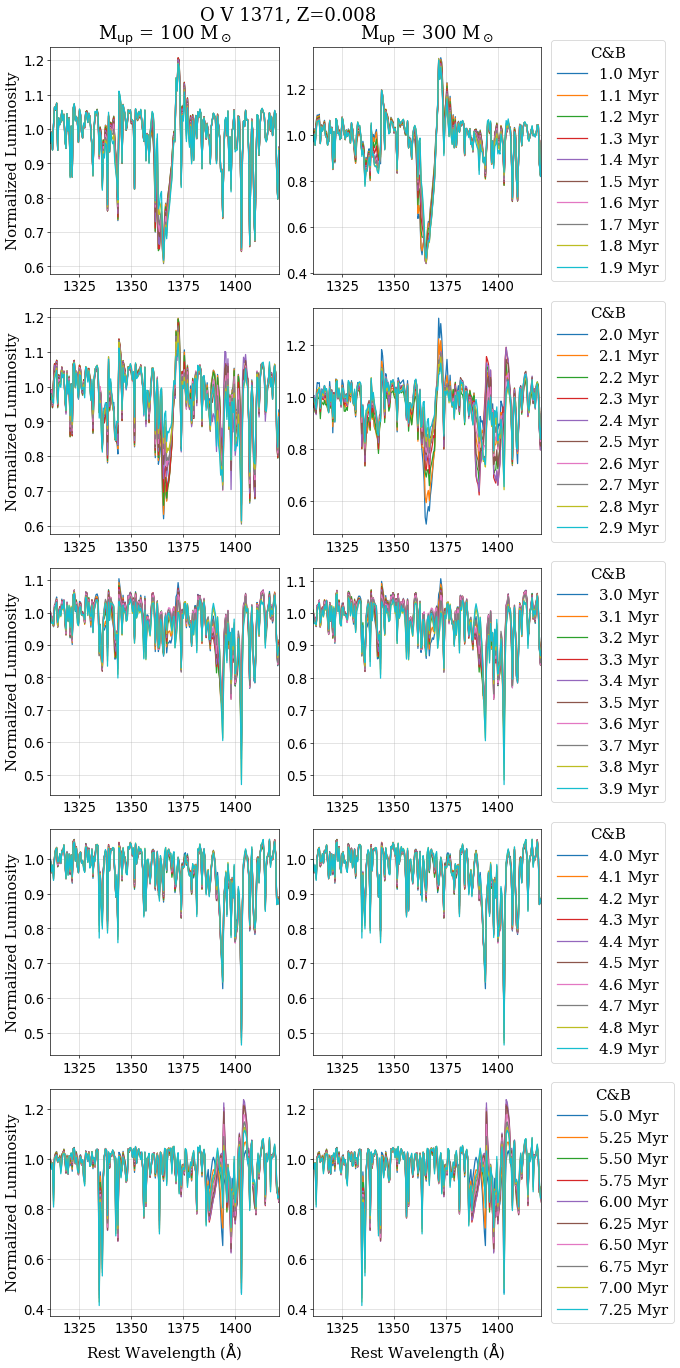}
    \caption{Similar to Figure~\ref{fig:cb19_1640} but for O\5 $\lambda$1371 and a metallicity of $Z=0.008$. The absorption component of the line is strong at ages $<3\,$Myr.}
    \label{fig:cb19_1371}
\end{figure}

Figure~\ref{fig:cb19_1371} is similar to Fig. \ref{fig:cb19_1640} but shows C\&B predictions for the O\5 $\lambda$1371 line profiles corresponding to $Z=0.008$, the metallicity of our best-fit model. The panels in the figure also include the Si\4 1400 doublet (which is strong when OB supergiants are present) and O\4 $\lambda$ 1340. Figure~\ref{fig:cb19_1371} shows that indeed O\5 is strong in absorption at ages younger than 3 Myr.  Note that the models predict that O\5 has a P-Cygni profile at the age of our best fit model but that in the observation, the line is only in absorption. This could be due to problems with the atmosphere models.

\begin{figure*}
    \centering
    \includegraphics[width=0.49\textwidth]
    {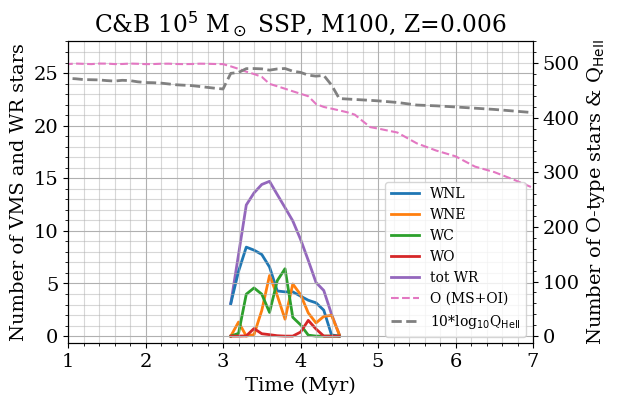}%
    \includegraphics[width=0.49\textwidth]
    {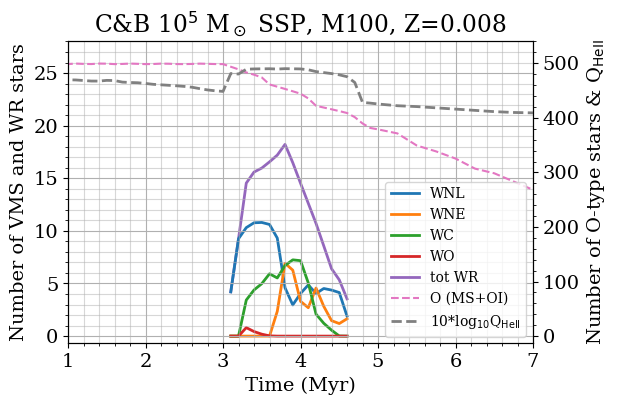}
    \includegraphics[width=0.49\textwidth]{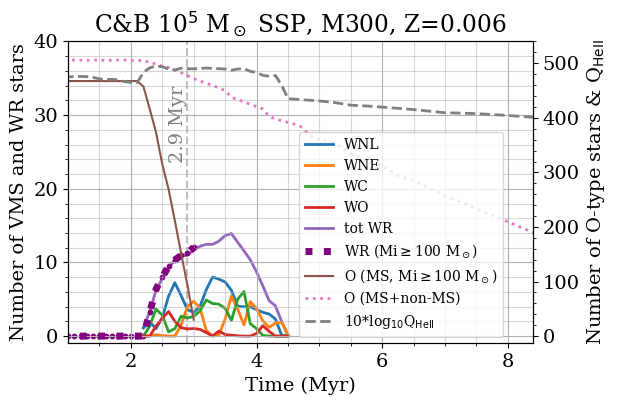}%
    \includegraphics[width=0.49\textwidth]{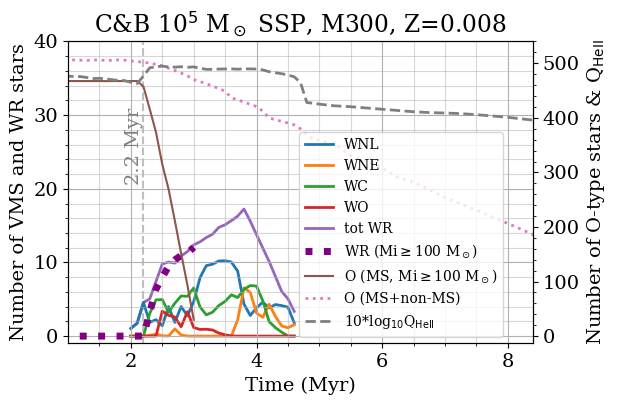}
    \caption{For C\&B SSP models of total initial mass equal to $10^5 \,M_\odot$ and all combinations of metallicity and upper mass limit of the IMF,  evolution of the numbers of stars of different types, as given by the legends. The different  solid curves are: blue (WNL), orange (WNE), red (WO), and purple (total number of WR stars). The purple squares on top of the purple curve correspond to WR stars whose progenitors have $\geq100\,M_\odot$. The rest of curves correspond to: MS O stars with initial masses, $M_i\geq100\,M_\odot$ (solid-brown); and O-type stars (MS + O I; dashed-pink). We include the value of 10 times the log of the He\2-ionising photon rate, Q$_{\rm HeII}$ in s$^{-1}$ (dashed-grey curve). Use the y-axis labels to know which y-axis to use. In the bottom panels, the vertical dotted lines give the ages of maximum He\2~1640 emission. Note that Q$_{\rm HeII}$ is boosted when the WR stars are present. As explained in \citealt{2023MNRAS.519.5656S}, this is unexpected for WN and WC stars.}
    \label{fig:num_stars}
\end{figure*}

 For all combinations of metallicity and upper mass limit of the IMF used in this work, Fig~\ref{fig:num_stars} shows the evolution of the number of massive stars of various types for a C\&B SSP with an initial mass in stars of 10$^5\,$M$_\odot$. The WR subtypes are defined in section 3.2 of \cite{2015MNRAS.452.1068C}. Additional details are provided in appendix A1 of \cite{2019MNRAS.490..978P}. For most star types given by the legend, the number of stars should be obtained from the left axis of the figure, whereas for O-type dwarfs + supergiants, which are more numerous, they should be obtained from the right axis. In the M100 models, which are shown in the top row of the figure, the WR stars are present at ages larger than 3 and less than 5 Myr, and stars of WNL subtype are present until slightly older ages at $Z=0.008$ relative to $Z=0.006$. In the M300 models, the WR stars that have progenitor masses of $\geq100\,M_\odot$ are marked with purple squares. WR stars appear earlier in the M300 models relative to the M100 models, and in the M300 models, earlier at the highest metallicity. In particular MS WNL stars are present at 2.2 Myr in the M300 model with $Z=0.008$, which is our preferred model. 

For completeness, Fig.~\ref{fig:num_stars} also shows the He$^+$ ionisation rate, Q$_{\rm HeII}$ (dashed-grey curve, use right axis). Note how Q$_{\rm HeII}$ goes up when the WNL stars (solid-blue curve) appear, and down when cWR stars have evolved. As discussed in \citealt{2023MNRAS.519.5656S}, the increase in Q$_{\rm HeII}$ during the WNL, WNE, and WC phases is unexpected due to the dense winds of these stars which trap the He$^+$ ionizing photons  \citep{2006A&A...449..711C, 2022arXiv221105424S}. The explanation provided in \citealt{2023MNRAS.519.5656S} is that C\&B use the PoWR atmosphere grids \citep{2015A&A...579A..75T}, which result in a set that can yield atmospheres with and without significant He\2-ionising flux; and that the C\&B population synthesis models tend to select weaker winded atmospheres that are more transparent to He\2-ionizing photons. Note that this is also the case for the population synthesis code Starburst99 \cite{1990ApJS...74...93R, 2010ApJS..189..309L}. The extent of the emission at $<228$\,\AA~is sensitive to the theoretical definitions of WNE (no H) and WNL (some H) versus empirical (WNE: high ionization, WNL low ionization) as well as the WNL threshold  The C\&B models use the WNL PoWR atmospheres described in the appendix of \cite{2019MNRAS.490..978P}.

For a discussion on the discrepancy between predictions for the number of WR stars from population synthesis codes Starburst99 \citep{2010ApJS..189..309L} and C\&B (models used in this work) relative to observations of these stars in the LMC, see the discussion in section 5 of \cite{2023MNRAS.519.5656S}. 

Figure 4 of \citet[][hereafter, S21]{2021MNRAS.503.6112S} shows a plot of the strength of stellar He\2 emission (equivalent width, EW) versus stellar C\4 absorption for a sample of extreme Wolf-Rayet (WR) galaxies and a selection of constant star formation population synthesis models. WR  galaxies are a subset of emission-line (or H\2) galaxies whose integrated spectra shows broad ($FWHM\sim10-20\,$\AA~or $640-1280$ km\,s$^{-1}$) He\2\,4686 emission \citep{Vacca1992}. In practice, the detection of a broad emission feature at $4650-4690$, which is attributed to WR stars and possibly includes additional emission lines (the so-called “blue bump”) is often simply used \citep{1999A&A...341..399S}. WR galaxies are distinguished spectroscopically from AGNs by their narrower Balmer lines and their forbidden line intensity ratios, which are indicative of stellar rather than non-stellar (i.e., power-law) excitation.

\begin{figure*}
    \centering
    \includegraphics[width=0.49\textwidth]{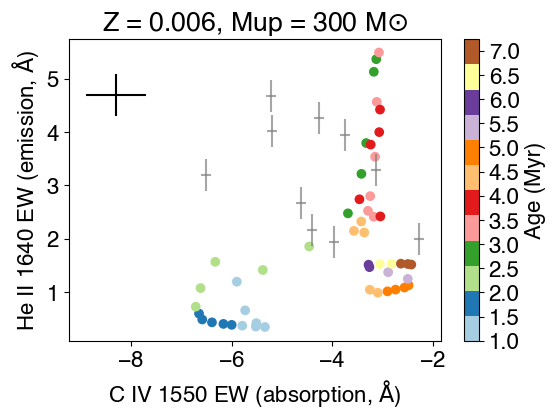}%
    \includegraphics[width=0.49\textwidth]{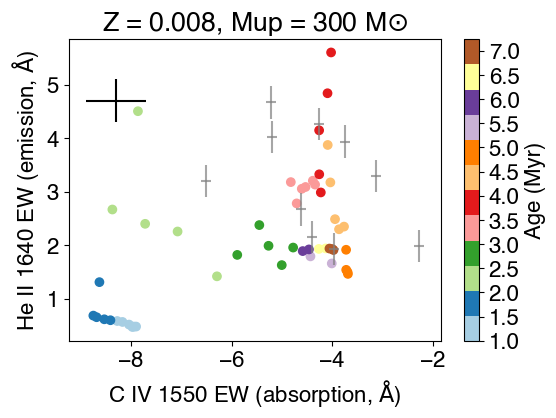}
    \caption{Similar to fig.~4 of S21 but for C\&B M300 SSPs instead of constant star formation, and only for the two metallicities used in this work. The left and right panels show the results for $Z=0.006$ and $Z=0.008$, respectively. The black cross represents the measurement errors in A1's observed EWs and the grey crosses are the approximate errors of the observations  presented in table 5 of S21 and corresponding to a sample of nearby star-forming galaxies with metallicities down to 12+log(O/H)=8.0. The filled circles are the model predictions at the ages given by the colour bar. Only the $Z=0.008$ model of 2.2 Myr approaches the EWs of A1. Also note the sensitivity of the C\4 strength to the metallicity.}
    \label{fig:fig5_ews}
\end{figure*}

The most extreme galaxy in the figure of S21 is not fully reproduced by the models they select. \nai~has a He\2 $\lambda1640$ equivalent width and metallicity that are comparable to that of their most extreme galaxy, SB 179 (their tables 4 and 5), but a larger C\4 1550 EW. In Fig.~\ref{fig:fig5_ews}, we show a plot similar to fig.~4 of S21 but for C\&B SSP models at the two metallicities used in the present work. The black cross is the A1. The grey crosses are the galaxies in table 5 of S21, which span a wider range in metallicity than our model metallicites and are only plotted to demonstrate that only one galaxy in S21 approaches the EW of A1's He\2 line. We use the same continuum and on-line windows for normalising the models and observation of A1, as in S21. Also following S21, we compute the EW values of the models and observation by dividing the He\2 emission and C\4 absorption components into rectangles (of width = 0.1 \AA, in our case) and summing the areas of the rectangles. In our case, we exclude the ISM components that contaminate the C\4 lines. The figure shows how the strength of the C\4 absorption depends on metallicity. Also note that only our best-fit model of 2.2. Myr approaches A1's EWs. In summary, VMS of $Z=0.008$ are the preferred explanation for the observed spectrum of A1.

\onlyinsubfile{

}

\onlyinsubfile{
\bibliographystyle{mnras}
\bibliography{REFERENCES/biblio}
}
A note could be added acknowledging that Fig. 11 is for LMC Z and some of the S21 galaxies have metallicities down to log O/H=8.0.

\onlyinsubfile{

\begin{figure*}
    \centering
    \includegraphics[width=0.5\textwidth]{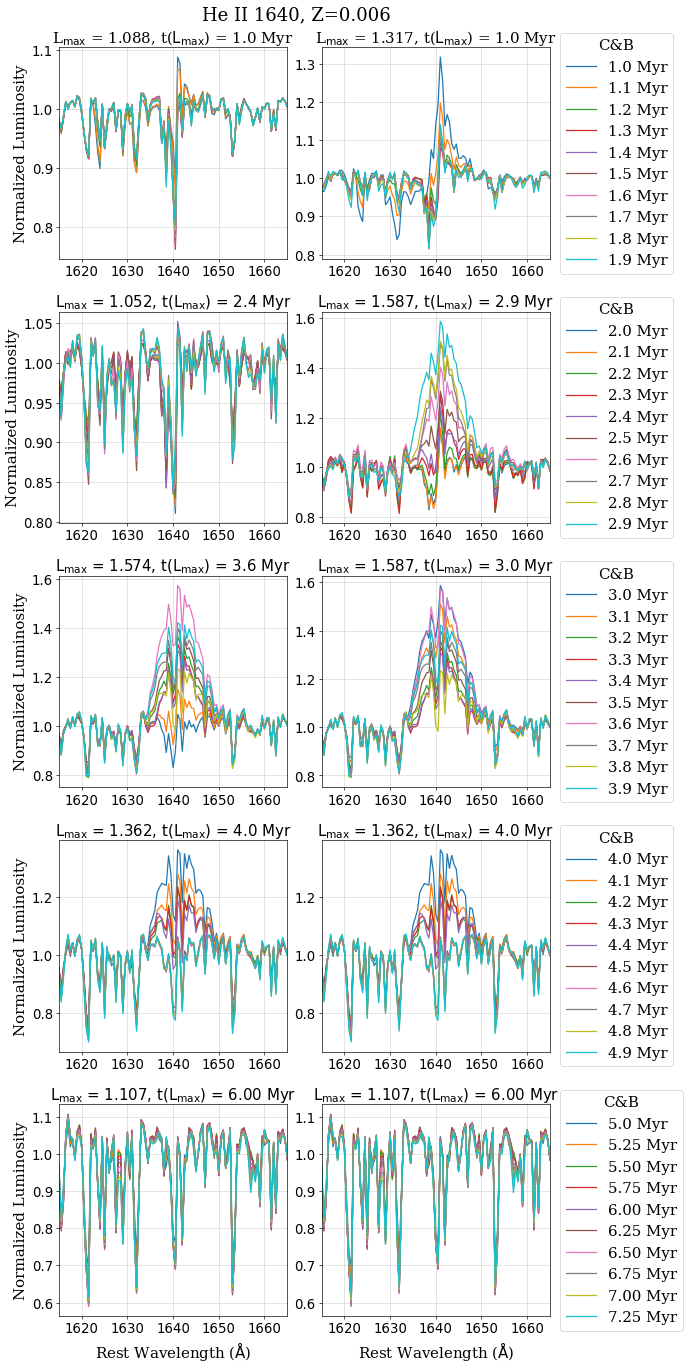}%
    \includegraphics[width=0.5\textwidth]{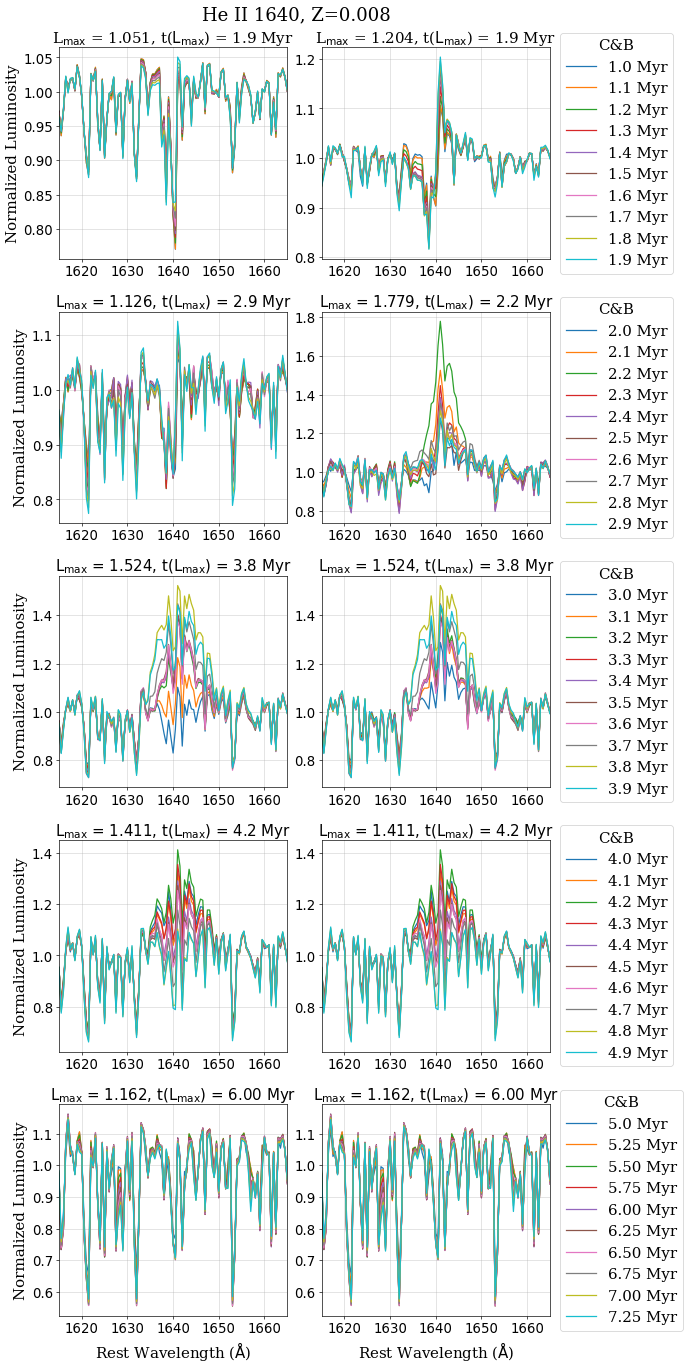}
    \caption{First two columns--. C\&B He\2$\,1640$ line-profile predictions for SSPs with $Z=0.006$ and $M_{\rm up}/M_\odot=100$ {\bf (1st column)} or $M_{\rm up}/M_\odot=300$ {\bf (2nd column)}. We use a linear fit between continuum windows [1615,\,1620] and [1660,\,1665] to normalise the He\2 lines. In each panel, the curves of different colours correspond to models of 10 different ages (except for the last row, which has 9), as given by the right-hand side legends. We give in each panel, the maximum luminosity in the wavelength range [1620,\,1660] $\AA$ that is reached in the panel, i.e., the amplitude of He\2 line when it is in emission; and the corresponding age. Models are available for ages in logarithmic space, which is why in the last row the time step changes and there are less curves. Last two columns--. Similar but for $Z=0.008$. The maximum amplitude of He\2 is obtained for $Z=0.008$, $M_{\rm up}/M_\odot=300$, and 2.2 Myr.}
    \label{fig:cb19_1640}
\end{figure*}

\bibliographystyle{mnras}
\bibliography{REFERENCES/biblio}
}

\subsection{A1´s properties.}\label{sec:5.2}

W14 estimate the present-day mass in stars of A1 from the ratio of the STIS luminosity at 1500\,\AA~(corrected for reddening and redshift), to that of an SSP model with an initial mass of $10^6\,M_\odot$, a \cite{2001MNRAS.322..231K} IMF, and an age of 3\,Myr. For the model, they use the empirical LMC/SMC stellar library. The luminosity at 1500\,\AA~is from the  continuum of the ensemble of stars.  For the above purpose, W14 use a distance of 11.5 Mpc to A1, colour excesses due to dust in the Milky Way (MW) and A1 of 0.08 and 0.013 mag, respectively, and the extinction laws of \citep{1999PASP..111...63F} and \citep{2003ApJ...594..279G} for the foreground (MW) and intrinsic (A1) reddening corrections, respectively. We adopt the latter parameters to obtain in a similar way luminosities from the COS observation (L$_{\rm 1500,obs}$) and the M100 and M300 models (L$_{\rm 1500, M100}$ and L$_{\rm 1500, M300}$, respectively). For obtaining the luminosities, we use the average flux over a $\pm3\AA$ wavelength interval around 1500\,\AA. In our case,  the models correspond to $Z=0.008$ and the age of highest amplitude of He\2\,$\lambda1640$, i.e., ages of 3.8 and 2.2 Myr, respectively. To a first approximation, the present-day mass in stars of A1 is given by the luminosity ratio. The luminosities and masses from W14 and this work are given in Tab~\ref{tab:a1_properties}. The masses from W14 and our M100 and M300 models are consistent with a SSC of $10^5\,M_\odot$. For a distance of 11.5 Mpc, the mass corresponding to our best-fit model is $1.9\times10^5\,M_\odot$, whereas it is $3.1\times10^5\,M_\odot$ when using the Virgo + GA + Shapley distance of 14.8 Mpc.

In addition, as in W14, we compute the present-day number of O stars in A1 ($N(O)$) by scaling the $N(O)$ values of our $10^6\,M_\odot$ SSP M100 and M300 models  by the above luminosity ratio, i.e., $N(O)_{\rm obs} = N(O)_{\rm theo} \times (L_{1500, theo} / L_{1500, obs})$. For M100, the $N(O)$ value is smaller by 19\% than in W14 because the age for M100 is 3.8 Myr instead of 3 Myr for W14. For our best-fit model and a distance of 11.5 Mpc, $N(O)_{\rm obs} =172$, which is about a factor of three times smaller than what we found in W14. It is 286 for a distance of 14.8 Mpc. 

In this work, we do not estimate the number of classical WR stars, since the presence of VMS indicates that classical WR stars have not appeared in the population yet.


\begin{table}
	\centering
	\begin{tabular}{l|l|l|c}
	\hline
	 Property & units & This work & W14  \\
  \hline
  \hfill & \multicolumn{3}{c}{Distance = 11.5 Mpc}\\
	\hline

     	L$_{\rm 1500, the, M100}$ & erg s$^{-1}$ & 1.9E+39 & 1.54E+39  \\
	L$_{\rm 1500, the, M300}$ & erg s$^{-1}$ & 2.9E+39& -   \\
      	L$_{\rm 1500,obs}$ & erg s$^{-1}$ & 5.4E+38 & 2.6E+38 \\
          	M$_{\rm *, obs, M100}$ & M$_{\odot}$ & 2.9E+05 & 1.7E+05  \\
	M$_{\rm *, obs, M300}$ & M$_{\odot}$ & 1.9E+05 & -  \\
N(O)$_{\rm the, M100}$ & stars & 1428 & 2793  \\
N(O)$_{\rm the, M300}$ & stars& 928 & -  \\
N(O)$_{\rm obs, M100}$ & stars & 408 & 467  \\
N(O)$_{\rm obs, M300}$ & stars & 172 & -  \\

  \hline
  \hfill & \multicolumn{3}{c}{{\bf Distance = 14.8 Mpc}}\\
	\hline
L$_{\rm 1500,obs}$ & erg s$^{-1}$ & 9E+38 & -  \\
M$_{\rm *, obs, M300}$ & M$_{\odot}$ & 3.1E+05 & -  \\
N(O)$_{\rm obs, M300}$ & stars & 286 & -  \\
\hline
	\end{tabular}
	\caption{Luminosity, mass in stars, and number of O stars for: model predictions corresponding to an SSP of $10^6\,M_\odot$ (properties with "the" in the subindex); observations of NGC3125-A1; and W14.}
	\label{tab:a1_properties}
\end{table}

\onlyinsubfile{

\begin{table}
    \centering
    \begin{tabular}{l|l|l|c}
    \hline
    Property & Units & Value & Reference \\
    \hline
         RA$_{\rm N3125}$ & h:m:s & 10:06:33.372 & 1\\ 
         Dec$_{\rm N3125}$ & d:m:s & -29:-56:-5.500 & 1 \\ 
         RA$_{\rm A1}$ & h:m:s & 10:06:33.280 & 2 \\
         Dec$_{\rm A1}$ & d:m:s & -29:-56:-6.800 & 2 \\
         Redshift & - & 0.003712 & 3 \\
         Distance 1 & Mpc & 11.5 & 4a \\
         Distance 2 & Mpc & $14.84\pm1.04$ & 4b\\
         $E(B-V)_{\rm{MW}}$ & mag & 0.073 & 5 \\
         $E(B-V)_{\rm{A1}}$ & mag & 0.130 & 6 \\
         12+log(O/H)$_{\rm A}$ & dex & $8.32\pm0.03$ & 7 \\
         \hline
    \end{tabular}
    \caption{Main properties of NGC~3125, region A, and SSC A1. Reference:\\
    (1) J2000 right ascension (h:m:s) or declination (d:m:s) of galaxy from Strasbourg astronomical Data Center (\url{http://cdsweb.u-strasbg.fr/}).\\
    (2) J2000 right ascension (h:m:s) or declination (d:m:s) of A1 used for COS observations from \url{https://archive.stsci.edu/missions-and-data/hst}.\\
    (3) Redshift from NASA/IPAC Extragalactic Database (NED, \url{https://ned.ipac.caltech.edu/}).\\
    (4a) Value derived from the Galactic Standard of Rest velocity using a Hubble constant of $H_0 = 75$\,km\,s$^{-1}$\,Mpc$^{-1}$, obtained by \citet{1999A&A...341..399S}, and adopted in W14 for comparison of the SSC masses and numbers of O and WR stars with previous works.\\
    (4b) Virgo + GA + Shapley distance \citep{2000ApJ...529..786M} from NED for cosmological parameters: $\Omega_{\rm{matter}}$ = 0.308, $\Omega_{\rm vacuum}$ = 0.692, and $H_0$ = 67.8 km\,s$^{-1}$\,Mpc$^{-1}$ \\
    (5) Colour excess due to dust in the Milky Way from: \url{https://noirlab.edu}.\\
    (6) Colour excess due to intrinsic dust from W14.\\
    (7) Oxygen abundance in the ionised gas from \citet{2006MNRAS.368.1822H}.    
    }\label{tab:ngc3125}
\end{table}

\onlyinsubfile{
\bibliographystyle{mnras}
\bibliography{REFERENCES/biblio}
}

\bibliographystyle{mnras}
\bibliography{REFERENCES/biblio}
}

\subsection{NGC~3125-A1 vs. a star-forming galaxy at $z\sim3$}\label{sec:5.3}

\begin{figure*}
    \centering
    \includegraphics[width=\textwidth]{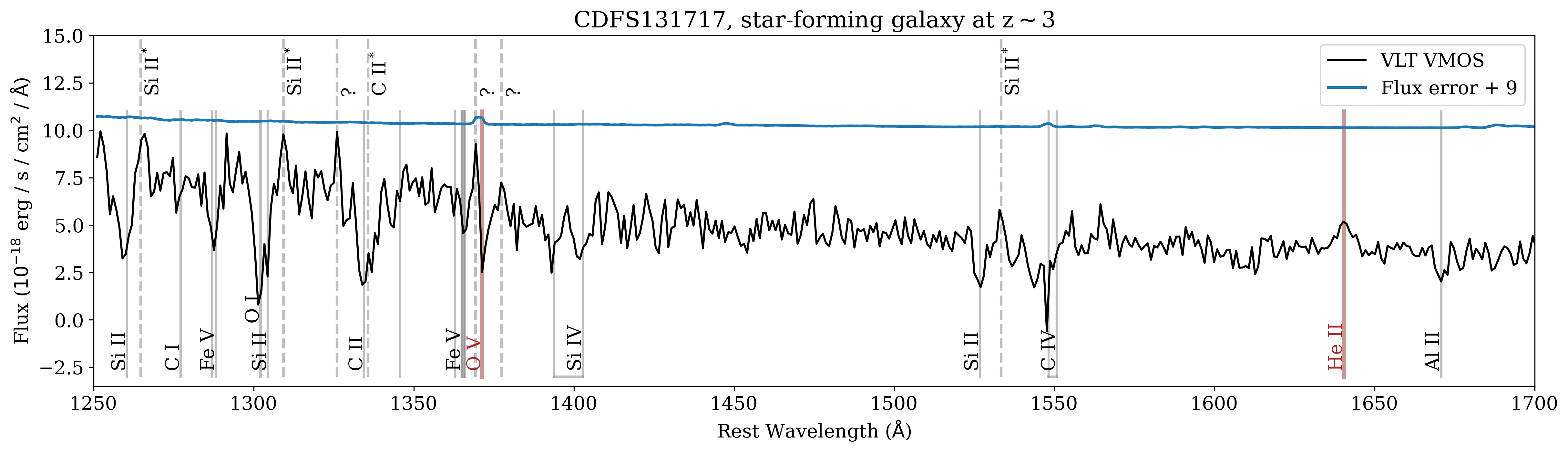}
    \caption{Reddening- and redshift-corrected \vlt~ VMOS spectrum of $z_{\rm spec}=3.071$ galaxy,
     CDFS131717 (solid-black curve), and corresponding flux errors (blue curve, offset for clarity). Based on \citet{2018Galax...6...63V}, we mark with labelled vertical lines the positions of: the He\2~1640 emission and tentative O\5~1371 absorption (solid brown); fine structure lines of Si\2$^*$ (1264.73, 1309.27, and 1533.43 \AA) and C\2$^*$ (1335.71 \AA; dashed grey), and other ISM and stellar features (solid grey).} 
    \label{fig:vandels}
\end{figure*}

In \S~\ref{sec:he2pb}, we mentioned high-redshift ($z=3.071$) and high star-forming galaxy, CDFS131717. This galaxy has a strong ($EW=2.5\pm0.1$\,\AA) broad ($FWHM\approx920\,$km\,$^{-1}$) He\2\,1640 emission line (Stanton et al. in prep). These properties were derived from a VLT VMOS spectrum, which has a resolving power of $R=580$, i.e., a resolution of $\sim508$ km s$^{-1}$ at 6500 \AA~near the observed-frame wavelength of the He\2~$\lambda1640$ line. The reddening- and redshif-corrected spectrum of the galaxy is shown in Figure~\ref{fig:vandels}.  Given that the He\2 line is the only broad UV line of CDFS131717, this object is unlikely to be an AGN. According to a full UV-optical SED fit, the galaxy has a stellar mass of $10^{9.7}\,M_\odot$  with a star-formation of $> 100\,M_\odot$/yr, making it highly star-forming for this redshift \citep[e.g.,][]{2014ApJS..214...15S}. The presence of broad He\2 emission with $EW > 2$\,\AA~also marks it out as rare within the general star-forming population ($< 2$\% of sources; \citealt{2020A&A...636A..47S}).

\onlyinsubfile{
\bibliographystyle{mnras}
\bibliography{REFERENCES/biblio}
}

 Figure~\ref{fig:vandels} shows a tentative detection of O\5\,1371 absorption. In VMS, this absorption is expected to be blueshifted (see Fig.~\ref{fig:cb19_1371}), but in CDFS131717, it appears to be at rest and next to an unidentified narrow emission line to the left of the absorption. Note that similar narrow emission lines are present at other wavelengths and correspond to Si\2 fluorescent emission, which is due to transitions to fine structure levels of the Si\2 ion. Similar lines are sometimes seen in galaxies (e.g., \citealt{2020ApJ...894..149W}). Also note that the flux error has a slight increase near the unidentified line. In the figure, the units of the flux and flux errors are the same but the flux errors are plotted offset for clarity. If the absorption line is O\5 1371, this could indicate the presence of VMS in the galaxy, particularly since it is accompanied by broad He\2 and strong C\4 lines. 

\begin{figure*}
    \centering
    \includegraphics[width=\textwidth]{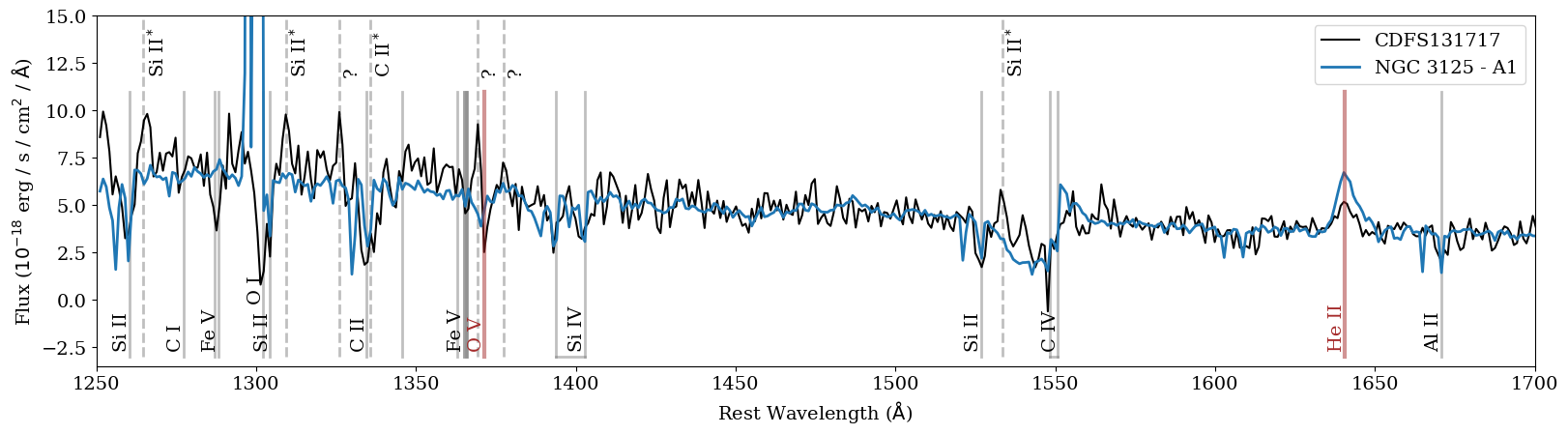}
    \caption{Comparison of the spectra of CDFS131717 (black curve) and \nai~(blue curve). Both spectra are corrected for reddening and redshift. A1's spectrum is degraded to the resolution of the VANDELS object and scaled to match its continuum flux at 1500\,\AA.}
    \label{fig:vandels_vs_a1}
\end{figure*}

In Fig.~\ref{fig:vandels_vs_a1}, we compare the shapes of the spectra of CDFS131717 (black curve) and \nai~(blue curve). A1's spectrum was degraded to match the resolution of the VANDELS object and scaled to match its flux in the continuum at 1500 \AA. Both spectra are corrected for reddening and redshift. An intrinsic slope of $\beta=-2.4$ (where $F_\lambda\propto\lambda^{\beta}$) was assumed for the continuum of CDFS131717. The O\5 absorptions almost coincide. On the other hand, the He\2 line of CDFS131717 is weaker. Note however that we are comparing a SSC (A1) to an entire galaxy, which has an underlying stellar continuum that is not present in A1.  

\begin{figure*}
    \centering
    \includegraphics[width=\textwidth]{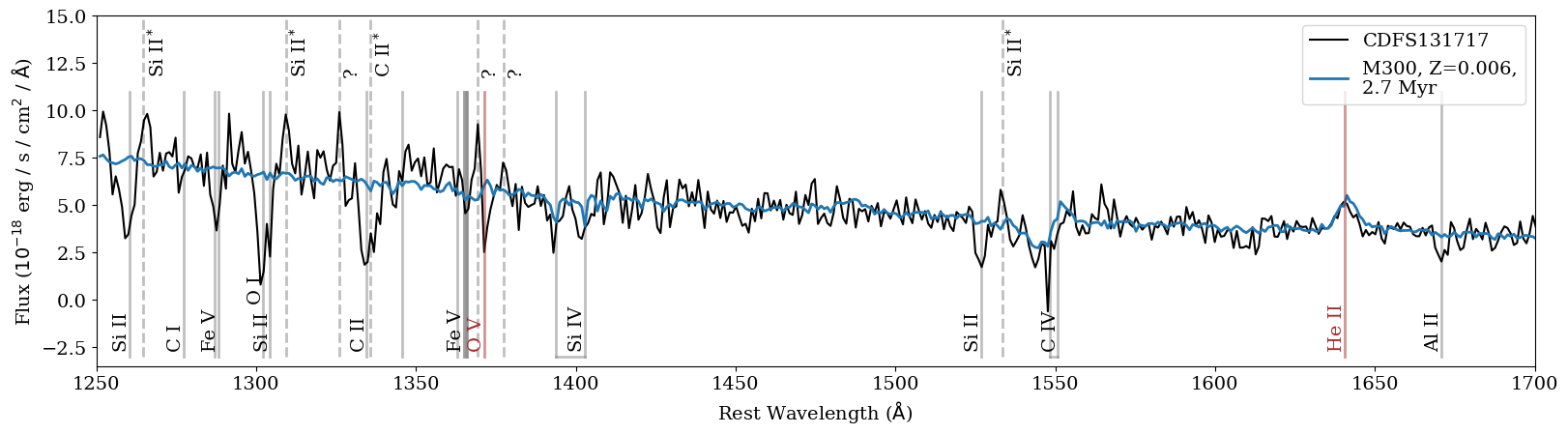}\\
    \includegraphics[width=\textwidth]{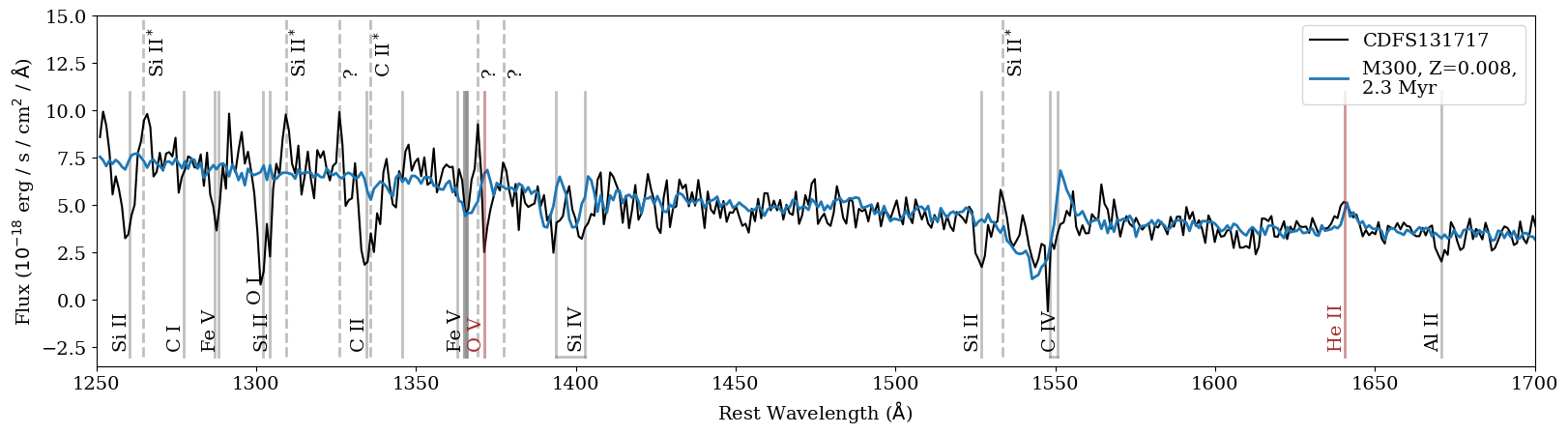}
    \caption{Similar to Fig.~\ref{fig:vandels_vs_a1} but we compare the spectrum of CDFS131717 (black curve) to M300 models with $Z=0.006$ and 2.7 Myr (top panel, blue curve); and $Z=0.008$ and 2.3 Myr (bottom panel). Although the fit to the C\4 and He\2 lines looks better in the top panel, at 2.7 Myr, the VMS are gone (see left panel of Fig.~\ref{fig:num_stars}), whereas at 2.3 Myr (bottom panel), they are still present (see right panel of Fig.~\ref{fig:num_stars}). However, in the bottom panel, the emission component of the predicted C\4 doublet is too strong.}
    \label{fig:vandels_vs_cb19}
\end{figure*}

In Fig.~\ref{fig:vandels_vs_cb19}, we compare the spectrum of CDFS131717 to M300 models (given the tentaitve detection of O\5 absorption). The lowest metallicity model (top panel, blue curve) seems to provide a reasonable fit to the stellar C\4 1550 and He\2~1640 features (the narrow C\4 ISM components of the observation are not accounted for in the model); but this model has weak O\5 in absorption. The highest metallicity model (bottom panel) approximately fits the He\2~line but produces too much C\4 1550 in emission. However, the latter model has stronger O\5 1371 in absorption. A fit to the galaxy with a more realistic (non-SSP) model is beyond the scope of our paper but will be presented in a more comprehensive study of CDFS131717 that also includes optical constraints (Stanton et al. in prep.). However, we note here that the tentative detection of O\5 absorption, clear detection of strong and broad He\2 emission, and rarity of high-redshift objects with similar He\2 line properties are all hints that short-lived VMS might be present in  CDFS131717. 


\onlyinsubfile{

\bibliographystyle{mnras}
\bibliography{REFERENCES/biblio}
}

\onlyinsubfile{
\bibliographystyle{mnras}
\bibliography{REFERENCES/biblio}
}

\subsection{Binaries}\label{sec:5.4}

There is now evidence that most massive stars are in close binary systems where the stars exchange mass and rotate, which affects their evolution relative to models where massive stars evolve as single non-rotating objects \citep{2012Sci...337..444S}. Figure \ref{fig:cb19_1640} shows that the 1640 emission fades after 5 Myr owing to the lack of WR stars from single stars beyond this age, but binaries will prolong the WR phase, as predicted by the work of \citealt{2018A&A...615A..78G} and observed by  \citealt{2021ApJ...912...16B}, who favour an age of approximately 10 Myr for Galactic cluster Westerlund 1, which is host to dozens of WR stars. 

The importance of including binaries in massive-star population studies is discussed in the review by  \cite{2022ARA&A..60..455E}. Several efforts have been carried out to determine the initial binary fraction, period distribution, and mass ratios of these stars \citep[e.g.][]{2012Sci...337..444S}. The effects of binaries are currently implemented in some spectral synthesis codes, e.g., BPASS \citep{2017PASA...34...58E} and the models of \cite{2019A&A...629A.134G}. 

\begin{figure}
    \centering
    \includegraphics[width=0.49\textwidth]
    {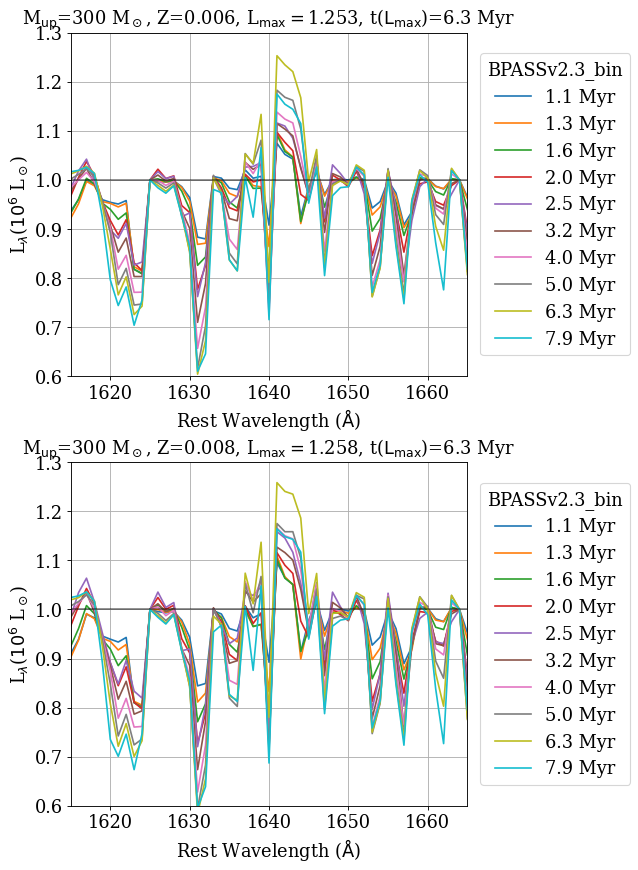}
    \caption{Evolution of He\2 1640 line profiles predicted by BPASS (version 2.3) M300 SSP models with close binaries, $\Delta$(log($\alpha$/Fe)) = 0, and $Z=0.006$ (top panel) and $Z=0.008$ (bottom panel), at the ages given by the legend. The models have a total initial mass in stars of $10^6$\,M$_\odot$. At the two metallicities, the maximum He\2 emission  occurs at $\sim6\,$Myr. At ages $\ge8$ Myr, this emission is weaker.}
    \label{fig:bpass1640}
\end{figure}

\begin{figure*}
    \centering
    \includegraphics[width=0.99\textwidth]{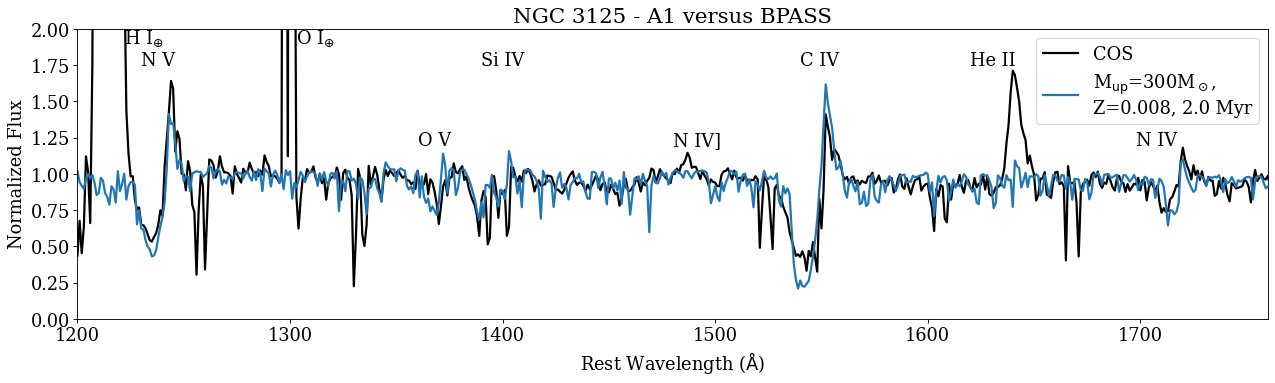}\\
    \includegraphics[width=0.99\textwidth]{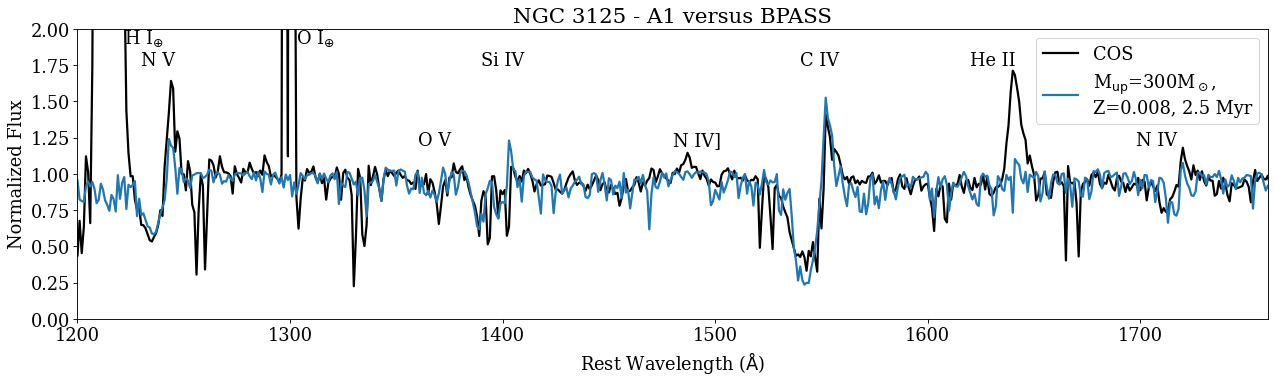}\\
    \includegraphics[width=0.99\textwidth]{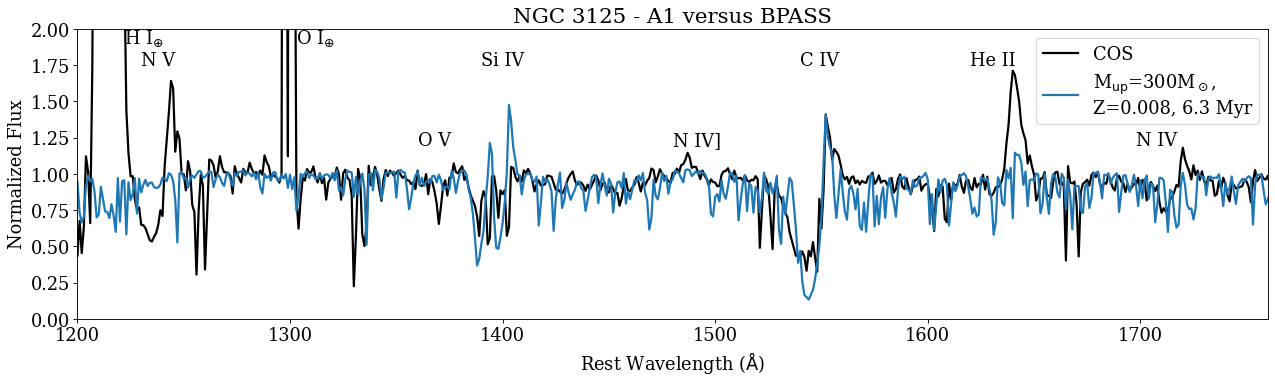}\\
    \caption{Comparison of BPASS M300 SSP models having close binaries and $Z=0.008$ (blue curves), with A1's observation degraded to the model's resolution and normalised (black curve). We show model at the ages given by the legend. The bottom panel is the case of maximum He\2 emission.}
    \label{fig:bpass_vs_a1}
\end{figure*}

The latest version of publicly available BPASS models (version 2.3) are presented in \citep{2022MNRAS.512.5329B}. In Fig.~\ref{fig:bpass1640}, we show BPASS v. 2.3 He\2 1640 line-profile predictions for M300 SSP populations with close binaries at the metallicities used in this work. We plot BPASS models with a standard mass ratio of $\alpha$-elements to iron, at all ages available between $\sim$1 to $\sim$8 Myr. At older ages, the strength of He\2 does not increase. In addition, BPASS models with the same total metallicity, Z, but in which mass ratio of $\alpha$-elements to iron has been modified by $\Delta$(log(alpha/Fe)) do not significantly change the strength of the He\2 profile. At the two metallicities shown in Fig.~\ref{fig:bpass1640}, the maximum strength of He\2 is reached at $6.3\,$Myr. Thus, binary models boost 1640 emission at higher ages with respect to single stars. In addition, at $6.3\,$Myr, the emission is slightly stronger for $Z=0.008$.  On the other hand, the maximum He\2 line strength is significantly less than that of A1. This can be seen in Fig.~\ref{fig:bpass_vs_a1}, where we compare BPASS models with $Z=0.008$ and different ages given by the legend. Ages 2.0 and 2.5 Myr are the closest available ages in BPASS to our best-fit C\&B model, which corresponds to 2.2 Myr. According to BPASS author J.J. Eldridge (private communication) some of the reasons that could explain the BPASS prediction for He\2 1640 are the following. At $Z=0.008$ the absorption lines in cooler stars start to get significant and decrease the strength of the He\2, as seen in Fig.~\ref{fig:bpass1640}. In addition, there are fewer VMS in the BPASS models at this metallicity because the models start having mass loss rates that are high enough to reduce the mass of VMS to more normal masses. Thus, the models have fewer stars near the Eddington limit so weaker He\2. Finally, the BPASS models do not include accretion-disk emission around stellar black holes and neutron stars, which might boost the He\2 emission. Going back to Fig.~\ref{fig:bpass_vs_a1}, also note that for none of the ages shown do the N\5 1240 and C\4 1550 features fit simultaneously. In summary, although BPASS models currently fail to reproduce strong stellar He\2 emission, population synthesis models should allow for modern mass-loss prescriptions, rotational mixing and close binary evolution.

\onlyinsubfile{
\bibliographystyle{mnras}
\bibliography{REFERENCES/biblio}
}

\section{Summary and conclusions}\label{sec:6}

In this paper, we present unpublished COS G160M observations of SSC \nai~that we analyse in combination with archival G130M spectra of the object. We focus on the modelling of the extreme He\2 emission of the SSC, which had represented a challenge for population synthesis models until recently \cite{2022A&A...659A.163M}.
\begin{enumerate}
    \item We determine that the He\2 line has a negligible contribution from nebular gas, is broad ($FWHM=1131\pm40$\,km\,s$^{-1}$), strong ($EW=4.6\pm0.5$\,\AA), and redshifted by $121\pm17$\,km\,s$^{-1}$ relative to ISM lines (Fig.~\ref{fig:he2_fit}). The broad line is the strongest ever detected in the nearby Universe, only comparable to one of the objects in the sample of nearby galaxies with broad He\2 emission that is analysed in \cite{2021MNRAS.503.6112S}.
    \item Although BPASS version 2.3 \citep{2022MNRAS.512.5329B} models with close binaries currently fail to reproduce strong stellar He\2 emission (Figs.~\ref{fig:bpass1640} and~\ref{fig:bpass_vs_a1}), it is clear that population synthesis models should allow for close binary evolution (\S~\ref{sec:5.4}). 
    \item On the other hand, a C\&B \citep{2019MNRAS.490..978P} SSP with single non-rotating stars, a metallicity of $Z=0.008$, and an age of 2.2\,Myr, where MS WR stars close to the Eddington limit are present, provides a very good fit to the He\2 line and other massive-star features in the wavelength range $\sim1150-1750$\AA~(Figs.~\ref{fig:cos_vs_cb19} and~\ref{fig:fig5_ews}). 
    \item We show that the presence of blueshifted O\5\,$\lambda1371$ absorption in the spectrum of a massive-star population provides an important clue of the youth of the population and possible presence of VMS. (Fig.~\ref{fig:cb19_1371}).
\end{enumerate}
We also present the spectrum of the broad-He\2 emitter, CDFS131717, which is an unlensed  strongly star-forming, low-metallicity galaxy located at $z\sim3.071$ (Fig.~\ref{fig:vandels_vs_cb19}). The tentative detection of O\5 absorption in the spectrum, clear detection of strong and broad He\2 emission, and rarity of the objects with similar properties at high-z are all hints that short-lived VMS might be present in  CDFS131717. A more detailed analysis of this galaxy accounting for the underlying stellar population and including constraints from the optical will be presented in Stanton et al. (in prep.).
Overall, our results show the effect of the improved formulation of stellar mass loss rates.  In conclusion, population synthesis models should include binaries, VMS, modern mass-loss prescriptions, and rotational mixing.
\onlyinsubfile{
\bibliographystyle{mnras}
\bibliography{REFERENCES/biblio}
}

\section*{Acknowledgements}


Based on observations made with the NASA/ESA Hubble Space Telescope, at the Space Telescope Science Institute, which is operated by the Association of Universities for Research in Astronomy, Inc., under NASA contract NAS5-26555. These observations are associated with programs \#12172 and \#15828.

A. Wofford and A. Sixtos acknowledge financial support from grant DGAPA/UNAM PAPIIT IN106922 (PI Wofford). G. Bruzual acknowledges financial support from UNAM grants DGAPA/PAPIIT IG100319 and BG100622. F. Cullen and T. M. Stanton acknowledge support from a UKRI Frontier Research Guarantee Grant (PI Cullen; grant reference EP/X021025/1).
 
We thank the referee for very helpful comments that improved the quality of this work. We also thank: J. Vink for pointing out that H-rich VMS have WR features, independent of He enrichment; P. Crowther and T. Shenar, who helped clarify why VMS and not classical WR stars are the main contributors to the UV spectrum of A1; J. C. Bouret and A. Sander, for helpful discussions about the use of the clumping factor in massive-star wind models; and E. Stanway and J.J. Eldridge for providing explanations for the weaker He\2 emission predicted by the BPASS models at $Z=0.008$.

\section*{Data Availability}
 
The \hst˜COS observations from PIDs 12172 and 15828 used in this work are available through the Barbara A. Mikulski archive for Space Telescopes at \url{https://mast.stsci.edu/search/ui/#/}. The VANDELS spectrum used in this work a publicly available and can be accessed via the ESO science portal (http://archive.eso.org/scienceportal/home). 



\bibliographystyle{mnras}
\bibliography{biblio}




\appendix

\section{WM-Basic and PoWR contributions}\label{appendix}

\begin{figure*}
    \centering
    \includegraphics[width=0.5\textwidth]{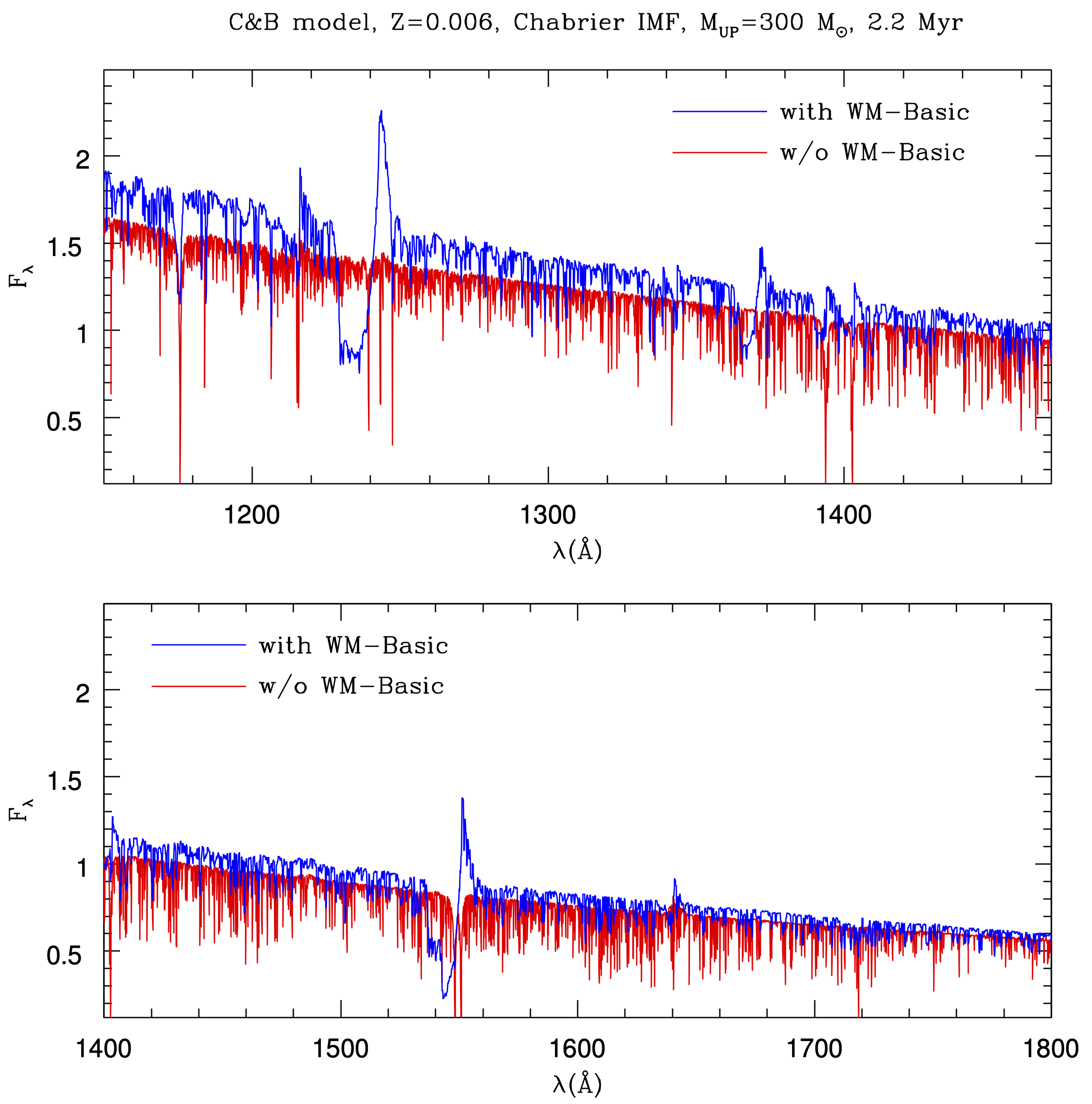}%
    \includegraphics[width=0.5\textwidth]{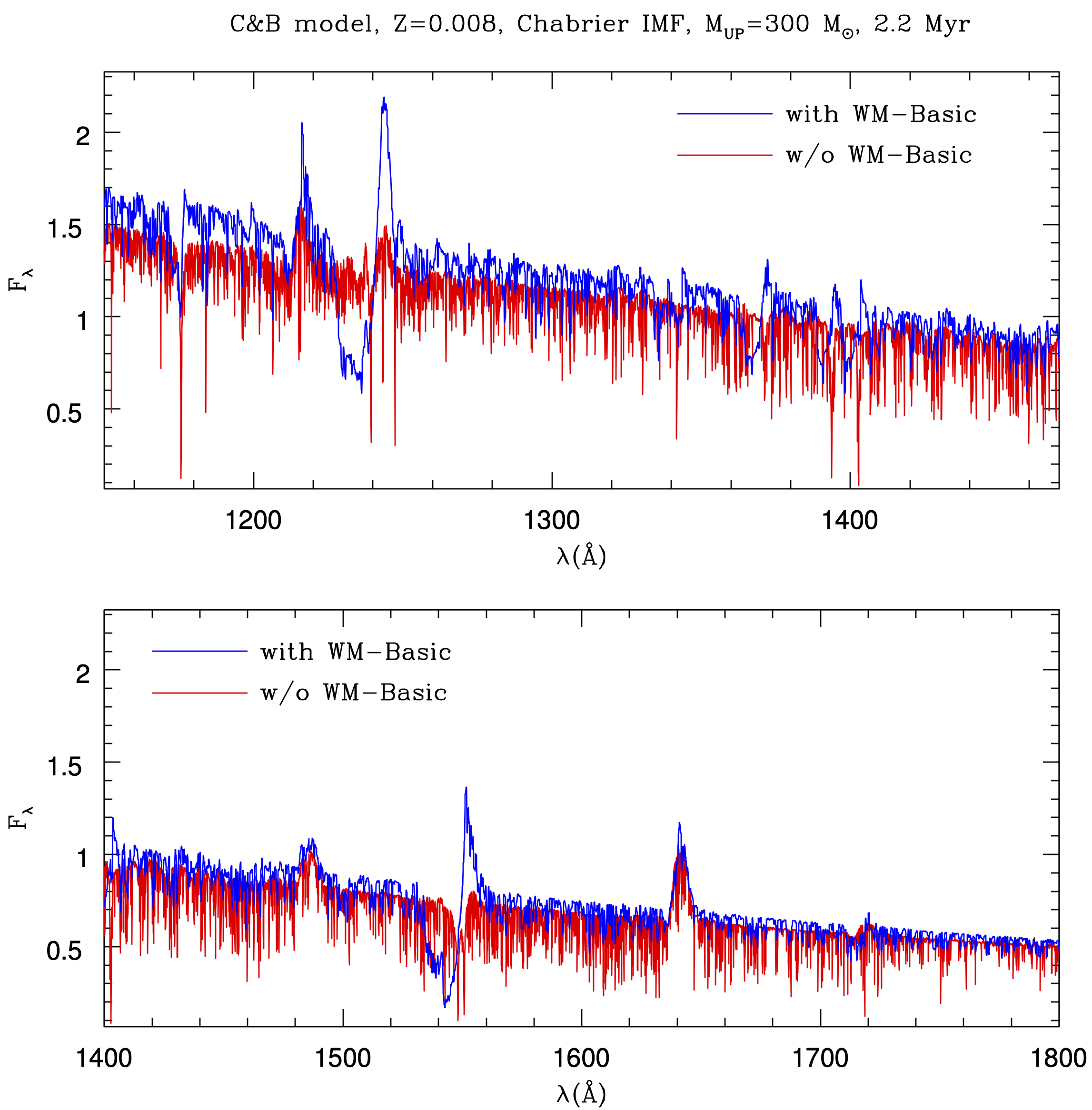}
    \caption{Left--. Difference in C\&B M300 SSP spectrum of $Z=0.006$ and 2.2 Myr, with (blue curve) and without (red-curve) the contribution of the WM-Basic models. The top and bottom rows show different wavelength ranges. At 2.2 Myr and $Z=0.006$, there are no WR stars of any kind in the population and a He\2 1640 P-Cygni like profile in only present when the WM-Basic models are included. Right--. Similar but for $Z=0.008$. At this metallicity, there are MS WR stars present. This is why in the bottom-right panel the He\2 1640 emission is dominated by the PoWR models.}
    \label{fig:appendix}
\end{figure*}

In Fig.~\ref{fig:appendix}, we show the effect on the final SED of not using the WM-Basic models, i.e., of  using plane parallel TLUSTY models with no stellar wind $+$ PoWR models (red curves). The blue curves  include Tlusty $+$ WM-Basic $+$ PoWR models. The top and bottom rows show different wavelength ranges. The figure illustrates that WM-Basic models dominate the P-Cygni like He\2 1640 profile of the population when no WR stars of any kind are present. This can be seen in the bottom-left panel, which corresponds to a C\&B M300 SSP of $Z=0.006$ and 2.2 Myr. On the other hand, the bottom-right panel of the figure, which corresponds to a similar model but with $Z=0.008$, shows that the He\2 1640 emission is dominated by the PoWR models. This is because at this metallicity, MS WR stars have already appeared at the same age, and in C\&B models, WR stars are modelled with PoWR. All this is in agreement with what is shown in the bottom panels of Fig.~\ref{fig:num_stars}, where the number of WNL stars as a function of time is plotted at the above two metallicities. At both metallicities, Fig.~\ref{fig:appendix} shows that WM-Basic models dominate the N\5 1240 and C\5 1550 P-Cygni profiles, which are due to the presence of O MS stars. Note that the model in blue shown in the bottom-right panel of Fig.~\ref{fig:appendix} is the one that best fits the COS observations of \nai. 

\bsp	
\label{lastpage}
\end{document}